\documentclass[onecolumn,letterpaper,12pt]{IEEEtran}% single spaced, for report
\usepackage{graphicx,url}
\usepackage{amsfonts,amsmath,amssymb}
\usepackage{amsmath}
\usepackage{amsfonts}
\usepackage{amssymb}
\usepackage{graphicx}%
\usepackage{fullpage}
\begin{document}

\title{Data Assimilation for Wildland Fires\\{\Large Ensemble Kalman filters in coupled atmosphere-surface models}}
\author{Jan Mandel, Jonathan D. Beezley, Janice L. Coen, and Minjeong Kim}
\maketitle
\thispagestyle{empty} 

\begin{center}
December 2007, Revised January 2009
\end{center}

%! comment out for report
%\CSMsetup \pagestyle{headings} 

\raggedbottom

Wildfires represent a growing and costly hazard to society, particularly as
homeowners build further into the wildland urban interface. Rather than a
simple tactical response to put out a reported ignition, the sheer magnitude
of areas engulfed in flames and intensity of the fires in fire prone areas
such as southern California call for a strategic approach with forecasting
tools to anticipate and mitigate fire behavior. Such tools do not currently exist.

A wildland fire is a complex multiscale process affected by nonlinear
scale-dependent interactions with other Earth processes. Physical processes
contributing to the fire occur over a wide range of scales. While the
characteristic scales of weather processes range over 5 orders of magnitude
from the hundred-km scale of large weather systems to the m-scale of
small-scale effects and eddies, the chemical reactions associated with the
thermal decomposition of fuel and combustion occur at scales of centimeters or
less to produce flamelengths up to $60$-m tall. Firelines travel with average
speeds on the order of a fraction of a m/s, while producing bursts of flame
that travel at 50 m/s, and chemical reactions occur on the order of seconds or
less. The wind and buoyancy produced by the fire are among the extremes of
atmospheric phenomena. Weather is the major external factor that affects fire
behavior, and two way interaction of fire and the atmosphere is essential --
fires are known to dramatically influence the weather in their vicinity. The
fire interacts with the atmosphere dynamics through fluxes of momentum, water
vapor, and heat, and with the soil through moisture and heat retention.

Data do not come as exact coefficients and initial and boundary conditions for
the model variables. Instead, various quantities only indirectly linked to the
model variables are measured at discrete points spread over time and space,
and the data are burdened with errors. Available data include fuel
distribution maps, gridded atmospheric state from larger-scale weather models
with assimilated weather data, point measurements from weather stations and
fire sensors, two-dimensional airborne infrared imagery of the fire, and maps
of fire extent produced by field personnel. The meteorological data are too
sparse to resolve mesometeorological ($2$--$20$-km scale) features.

%\subsection{Control issues and challenges}

%\subsubsection{A quick introduction what data assimilation is}

A computational model can capture only a select fraction of the significant
mechanisms in the wildland fire process. Even if an accurate model existed,
the data are not complete and accurate enough to make an accurate prediction
possible. Also, one challenge of modeling is to estimate the accuracy of a
forecast; a forecast has little value without additional information on what
confidence level may be placed on it. Therefore, it is natural to consider
statistics-based data assimilation methods. These methods include parameter
and state estimation. Then the state of the model is the probability
distribution of possible wildfire scenarios and the data are entered as data
likelihood, which combines the information about the values of the
measurements and the probability distribution of measurement errors. The data
assimilation methods considered here proceed in analysis cycles. In each
cycle, the model state is advanced in time, then new data are injected at the
end of the cycle by combining the probability distribution of the state with
the data likelihood.

%\subsubsection{Fire properties that make it challenging}

However, the evolution of fire is highly nonlinear and the ignition is sharp or
even discontinuous on the model scale. Statistical variability in additive
corrections to the state may cause spurious ignitions, and additive
corrections are not adequate to make changes to the location of the fire.
Probability distribution of the fire state can be multimodal and centered
around the burning and not burning states at any given point in space. The
whole fire state may concentrate around more than one distinct scenario, such
as whether or not the fire jumps a road.

%\subsubsection{Why are classical methods unsuitable}

Classical data assimilation methods, such as statistical interpolation, 3DVAR,
and 4DVAR~\cite{Kalnay-AMD-2003} represent probability distributions only by
the mean and the covariance. But only the normal distribution is completely
determined by the mean and the covariance, so the strong non-Gaussian
character of fire makes the use of those methods questionable. The
complexities of the model and sharp nonlinearities rule out adjoint methods,
such as 3DVAR and 4DVAR, which are developed for atmospheric models, where the
adjoint code can be produced by reversing the order of certain loops. Another
option are ensemble methods. Although an ensemble can, in principle, represent
a strongly non-Gaussian distribution, the EnKF \cite{Evensen-2007-DAE}
formulas are based on the assumption that the ensemble is a sample from a
Gaussian distribution. Particle filters \cite{Doucet-2001-SMC} also use an
ensemble, and do not make the Gaussian assumption. However, particle filters
are known to require a large ensemble size, which grows fast for
high-dimensional problems. This requirement makes them generally unsuitable
for problems with more than a handful of variables, in particular for fire
models, which have millions of gridpoints, with several model variables
associated with each gridpoint.

%\subsubsection{Synopsis: new methods that work}

The goal of this article is to develop nontraditional ensemble Kalman filter
methods for the forecast of the behavior of wildland fires. The ultimate goal
is to estimate the state of the fire from available data and deliver a
short-term prediction faster than real time taking into account the interactions
between the fire and the atmosphere. Although fire is a complicated process,
simple deterministic models that run fast can capture its essential behavior.
Two such models are presented, a reaction-diffusion model based on simplified
chemical kinetics \cite{Mandel-2008-WFM}, and a new level-set implementation
of a semi-empirical fire propagation model \cite{Clark-2004-DCA} coupled with
the Weather Research and Forecasting (WRF) atmospheric model \cite{WRF}. The fire models
are combined with two enhanced versions of the EnKF. The \emph{regularized
EnKF}\/~\cite{Johns-2008-CEK} incorporates an assumption that the physical
fields are (mostly) smooth, thus helping to prevent large variations in
temperature and spurious ignition. The \emph{morphing EnKF}
\cite{Beezley-2008-MEK} exploits methods of image processing to combine
spatial and amplitude corrections to the state.
%The \emph{predictor-corrector filters}
%combine EnKF and particle filters to handle strongly non-Gaussian
%distributions with small ensembles.

\section{How Should Wildland Fires Be Modeled?}

In order to be faster than real time, fire models need to strike a balance
between fidelity and fast execution. This article considers only two
simplified 2D models for a fire in a layer just above the ground; see
\textquotedblleft Wildland Fire Modeling\textquotedblright\ for other
approaches. Both models here are posed in the horizontal $\left(  x_{1}%
,x_{2}\right)  $ plane, have a single vertical layer, and can be coupled with
the atmosphere using the same variables. Their state variables and equations,
however, are different.

The first model consists of the differential equations of chemical kinetics
with two physical variables, the fuel temperature and the fuel mass. The speed
of propagation of the fire is a property of the solution of the system of
differential equations and thus it is determined by the coefficients of the
equations only indirectly. In spite of its simplicity, this model captures
naturally a number of fire phenomena, such as fire jumping a fuel break due to
heat transfer across the break or fire being extinguished by increasing the
heat outflow. On the other hand, the coefficients of the model are hard to
calibrate to the observed fire behavior and the model requires a very fine
grid and a very small time step to resolve the nonlinear effects in the
leading edge of the fire, which is essential for correct fire propagation.

The second model is semi-empirical and it postulates the fire propagation
speed normal to the fireline as a function of wind and terrain slope, and
assumes an exponential decay of fuel from the time of ignition. Thus, the fire
propagation speed can be input directly from known fire behavior for various
fuels. The model can run on a relatively coarse grid, with the mesh step
dictated only by the desired resolution of the shape of the fire region. The
model has been validated on laboratory data and a number of forest fires.
However, the model does not capture gradual changes in fire behavior, which
would have to be implemented by ad-hoc modifications. The temperature or the
fire intensity are not variables of this model and the heat produced is only a
function of the time from ignition and of the fuel properties.

Although both models can be in principle coupled with an atmospheric model, we
have implemented the coupling only for the semi-empirical model. These two models are
representatives of the physics-based and semi-empirical 
approaches in wildfire modeling, respectively. Although a coupling of the two
fire models could be considered, this was not done in the work presented, and
either one or the other are used separately.

\subsection{Reaction-diffusion equations model}

Consider a model consisting of a layer of fuel with concentration $F=F\left(
x,t\right)  $ ($kg/m^{2}$), $x=\left(  x_{1},x_{2}\right)  $ (m), and
temperature $T=T\left(  x,t\right)  $ (K). The fuel is assumed to burn at the
relative rate $r\left(  T\right)  $ $\left(  \text{1/s}\right)  $, dependent
only on the temperature. The balance of heat in the fuel layer is described by
the partial differential equation
\begin{equation}
c\frac{\partial T}{\partial t}=\nabla\cdot\left(  k\nabla T\right)
-(\overrightarrow{v}+\gamma\nabla z)\cdot\nabla T+\underbrace{AF\,r\left(
T\right)  -\overbrace{C\left(  T-T_{a}\right)  }^{H}}_{h\left(  T\right)  },
\label{eq:heat}%
\end{equation}
where $\nabla=\left(  \frac{\partial}{\partial x_{1}},\frac{\partial}{\partial
x_{2}}\right)  $ is the gradient, $\cdot$ is the dot product, $c$, $k$, $A$,
and $C$ are coefficients generally dependent on $x$ and $y$, $\overrightarrow
{v}=\overrightarrow{v}\left(  x,t\right)  $ is the wind speed, and
$\gamma\nabla z$ is the correction for the gradient of the terrain elevatin
$z$ (since fire moves faster uphill). The term $c\frac{dT}{dt}$ is the heat
flux (J/m$^{2}$s) which is absorbed in the fuel and changes the fuel layer's
temperature. The diffusion term $\nabla\cdot\left(  k\nabla T\right)  $ models
the heat transfer by short range radiation and air mixing, which causes the
fire to spread between neighboring fuel particles (such as twigs and
branches). The advection term $\overrightarrow{v}\cdot\nabla T$ is the heat
flux due to wind. The external heat flux $h\left(  T\right)  $ consists of the
heat flux $AFr\left(  T\right)  $, generated by burning the fuel, minus the
heat flux%
\begin{equation}
H=C\left(  T-T_{a}\right)  , \label{eq:heat-out-pde}
\end{equation}
escaping to the atmosphere at the ambient temperature $T_{a}$ according to the
Newton's law of cooling. The change in the fuel supply due to the burning is
described by the fuel balance equation%
\begin{equation}
\frac{dF}{dt}=-F\,r(T), \label{eq:fuel}%
\end{equation}
which is a collection of independent ordinary differential equations
associated with the surface points $\left(  x_{1},x_{2}\right)  $.\ The
reaction rate is taken to be%
\begin{equation}
r\left(  T\right)  =De^{-B/(T-T_{a})}, \label{eq:reaction}%
\end{equation}
where $B$ and $D$ are coefficients. The reaction rate (\ref{eq:reaction}) has
the form of the Arrhenius reaction rate from chemical kinetics except for the
offset $T_{a}$, which is to guarantee that the reaction rate is zero at the
ambient temperature. Although the model (\ref{eq:heat})--(\ref{eq:reaction})
is simple, its solutions exhibit qualitative behaviors characteristic of
combustion, known from chemical kinetics and the theory of reaction-diffusion
equations \cite{Frank-Kamenetskii-1955-DHE,Grindrod-1996-TAR}.

\begin{figure}[ptb]
\begin{center}
\includegraphics[width=3in]{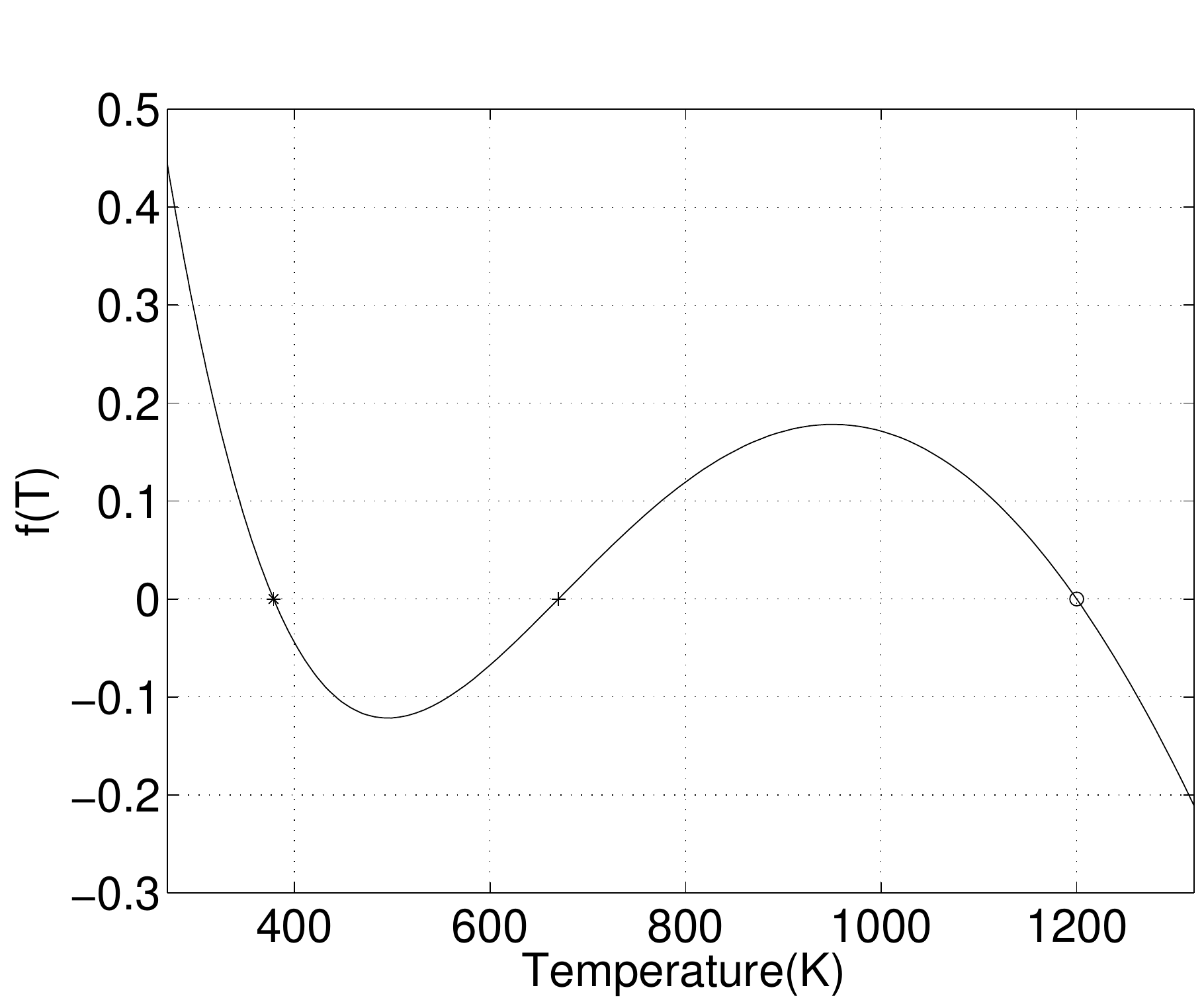} \newline
\end{center}
\caption{An example of reaction heat balance function $h(T)$ from modified
Arrhenius reaction dynamics with cooling (\ref{eq:heat}). The three zeros of
the function, in increasing order, are the ambient temperature $T_{a}$, the
auto-ignition temperature $T_{i}$, and the high burning temperature $T_{c}$.
For the temperature $T$ between $T_{a}$ and $T_{i}$, the heat produced by the
reaction is less than the heat loss by cooling. Once the $T$ higher than the
$T_{i}$, but less than $T_{c}$, the reaction heat is more than the heat loss
by cooling, and fire results. Thus, $T_{i}$ is unstable equilibrium. Since the
heat loss by cooling grows linearly with the temperature but the reaction rate
is bounded, for high enough temperature, $T>T_{c}$, the heat loss by cooling
prevails. Reproduced from \cite{Mandel-2008-WFM}.}%
\label{fig:f}%
\end{figure}

Neglecting for the moment the change in the fuel supply $F$, (\ref{eq:heat})
has three equilibria, which are given by the zeros of the function $h\left(
T\right)  $ (Fig. \ref{fig:f}). The smallest zero is the ambient temperature,
where there is no reaction and no heat flux to the atmosphere. This smallest
zero is a stable equilibrium because the derivative $h^{\prime}\left(
T_{a}\right)  <0$. The middle zero is the ignition temperature $T_{i}$, which
is an unstable equilibrium, $h^{\prime}\left(  T_{i}\right)  >0$. At the
ignition temperature, the heat generated by the reaction balances the heat
escaping to the atmosphere, but a small increase of the temperature above
$T_{i}$ creates a positive heat flux, which increases the temperature further,
thus leading to the development of fire. The temperature climbs toward the
largest zero of $h$, but then decreases as the fuel burns off following
(\ref{eq:fuel}).

Solutions of the system (\ref{eq:heat})--(\ref{eq:reaction}) exhibit
propagating combustion waves, at least for coefficients in certain ranges. At
the leading edge of the combustion wave, the diffusion term spreads the heat
into not-yet-ignited fuel and increases its temperature. Once the temperature
is high enough for the heat generated by the reaction to overcome the heat
loss, the temperature increases quickly and we say that the fuel ignites.  
The fuel then
burns off and cools down, creating the trailing edge of the combustion wave.
As a result, the combustion wave moves into yet-unburned fuel and leaves
behind fuel burned down to a quantity no longer sufficient to sustain the reaction.

\begin{figure}[ptb]
\begin{center}
\includegraphics[width=4in]{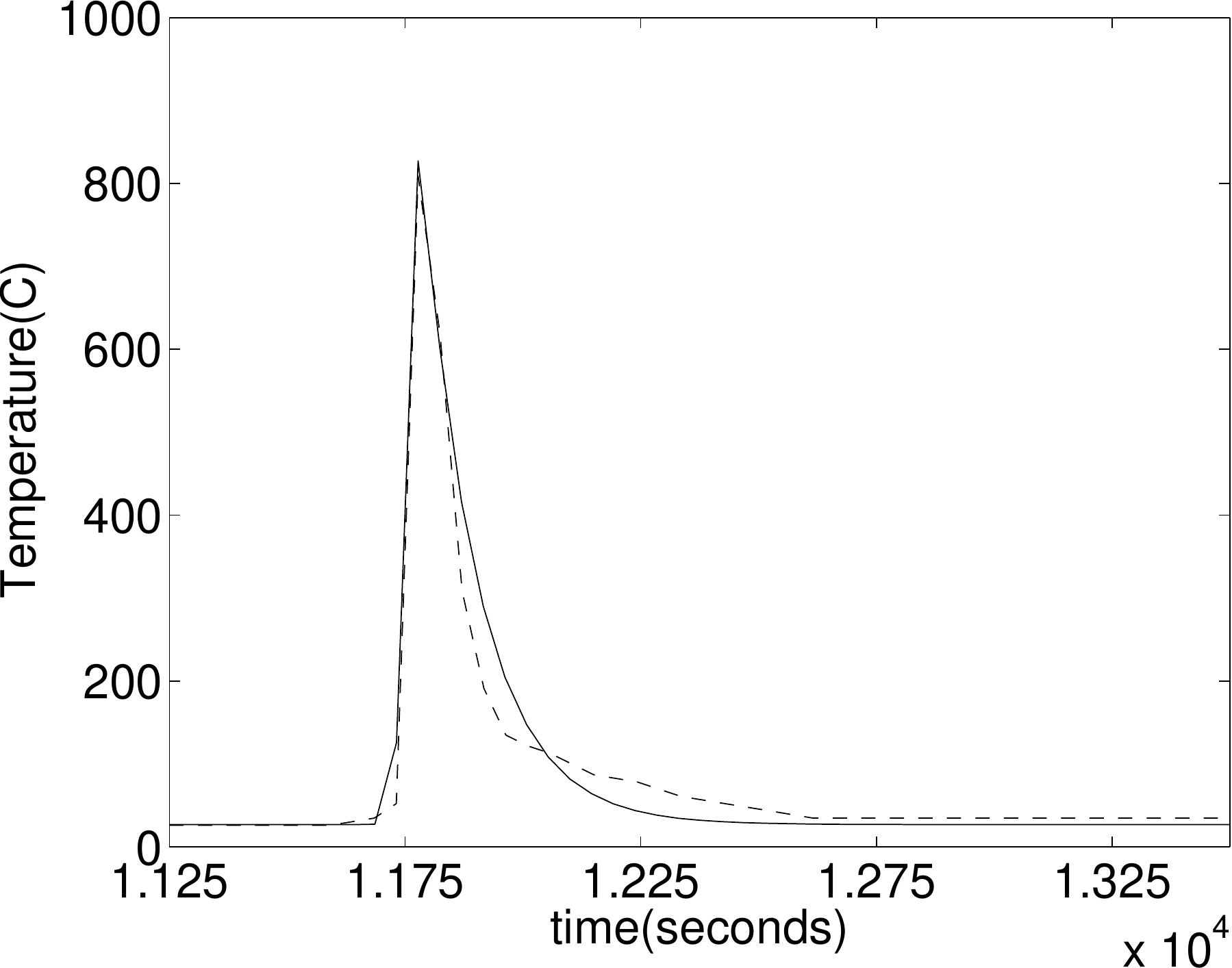} \newline
\end{center}
\caption{Measured and computed temperature as the fire passes over a sensor in
a fixed location. The measured time-temperature profile (dash line) in a grass
wildland fire is digitized from \cite{Kremens-2003-MTT}. The computed computed
profile (solid line) is obtained from a simulation by a numerical solution of
the reaction-diffusion equations (\ref{eq:heat}) -- (\ref{eq:reaction}). The
model was ignited by raising the temperature in a subinterval, then two waves
develop traveling away from the ignition and eventually stabilize. The
computed profile shows such wave, traveling to the left. Reproduced from
\cite{Mandel-2008-WFM}.}%
\label{fig:tp}%
\end{figure}

Unfortunately, taking coefficients from the physical and chemical properties of the fuel and expecting to obtain the
correct combustion wave speed is hopeless. The reason is that coefficients are
a homogenized form of unknown microscale data and many important physical
processes are not modeled at all. For example, the model ignores the effect of
the disappearance of fuel on the total heat capacity $c$ of the fuel layer,
storage of the heat in the ground, the chemical kinetics of intermediate
products of combustion, the fine scale fire dynamics on the surface and in the
interior of the wood, and evaporation of moisture. Fortunately, suitable
coefficients can be identified from observed macroscopic behavior of the fire.
In \cite{Mandel-2008-WFM}, the coefficients are determined from temperature
profile measured by a stationary sensor passed over by a moving combustion
wave, and from the measured speed of the combustion wave (Fig. \ref{fig:tp}).
Combustion wave speed is also called the fire spread rate.

The model (\ref{eq:heat})--(\ref{eq:reaction}) is discretized by standard
finite difference techniques, with a~sufficiently fine mesh to resolve the
ignition region and upwinding of the advection term for stability. Ignition is
achieved by an initial condition with a temperature well over the ignition
temperature in a given ignition region, at least several meshcells large.

The computation visualized in Fig. \ref{fig:tp} was done with mesh step size
$2$ m on an interval $1000$ m long and time step $0.01$ s. A\ realistic fire
model may require a domain size $10$ km by $10$ km, resulting in $50$ million
variables. Mosaic character of fire makes the use of local mesh refinement
difficult. Some fuels, such as grass, exhibit a reaction region only cm wide,
thus requiring a mesh step of the order of $0.1$ m or less, which would result in models
with more than $10^{11}$ variables.

\subsection{Fireline propagation model}

The semi-empirical fire propagation model imposes the fire spread rate
directly, replaces the leading edge of the combustion wave by instantaneous
ignition, and replaces the fuel depletion rate by an imposed one.

Consider fire burning in the area $\Omega=\Omega\left(  t\right)  $ in the
$\left(  x_{1},x_{2}\right)  $ plane, with the boundary $\Gamma=\Gamma\left(
t\right)  $, called the fireline, and with outside normal $\overrightarrow
{n}=\overrightarrow{n}(x,t)$, $x\in$ $\Gamma\left(  t\right)  $. The time of
ignition $t_{i}\left(  x\right)  $ at a point $x=\left(  x_{1},x_{2}\right)
\in$ $\Omega\left(  t\right)  $ is defined as the time when the point is at
the fireline, $x\in\Gamma\left(  t_{i}\left(  x\right)  \right)  $.

The model postulates that the fireline evolves with a given spread rate
$S=S\left(  x,t\right)  $ in the normal direction. The spread rate is a
function of the is the components of the wind $\overrightarrow{v}$ and the
terrain gradient $\nabla z$ in normal direction $\overrightarrow{n}$ to the
fireline,
\[
S=S\left(  \overrightarrow{v}\cdot\overrightarrow{n},\nabla z\cdot
\overrightarrow{n}\right)
\]
as \cite{Clark-2004-DCA,Rothermel-1972-MMP}
\begin{equation}
S=\left\{
\begin{tabular}
[c]{ll}%
$0,$ & if $\widetilde{S}<0,$\\
$S_{\max},$ & if $\widetilde{S}>S_{\max},$\\
$\widetilde{S},$ & otherwise,
\end{tabular}
\right.  \widetilde{S}=\min\left\{  B_{0},R_{0}+\phi_{W}+\phi_{S}\right\}  ,
\label{eq:Rothermel}%
\end{equation}
where $R_{0}$ is the spread rate in the absence of wind, $\phi_{W}=a\left(
\overrightarrow{v}\cdot\overrightarrow{n}\right)  ^{b}$is the wind correction,
$\phi_{S}=d\nabla z\cdot\overrightarrow{n}$ is the terrain correction, $a$,
$b$, and $d$ are given coefficients, and $B_{0}$ is the backing rate, that is
the minimal fire spread rate even against the wind. A small backing rate of
spread must be specified, since fires are known to creep upwind on their
upwind edge due to radiation.

In the burning area, the model postulates that the fuel decreases
exponentially from the ignition time,%
\begin{equation}
F(x,t)=\left\{
\begin{tabular}
[c]{ll}%
$F_{0}\left(  x\right)  e^{-\left(  t-t_{i}\left(  x\right)  \right)
/W\left(  x\right)  },$ & if $x\in\Omega\left(  t\right)  $,\\
$F_{0}\left(  x\right)  ,$ & otherwise,
\end{tabular}
\ \ \ \ \ \right.  \label{eq:burning}%
\end{equation}
where $F_{0}\left(  x\right)  $ is the initial fuel supply and $W\left(
x\right)  $ is the time constant of the fuel.\ The heat flux from the fire to
the atmosphere is given by the amount of fuel burned,%
\begin{equation}
H=-A\left(  x\right)  \frac{d}{dt}F\left(  x,t\right)  =\left\{
\begin{tabular}
[c]{ll}%
$A\left(  x\right)  F_{0}\left(  x\right)  \frac{e^{-\left(  t-t_{i}\left(
x\right)  \right)  /W\left(  x\right)  }}{W\left(  x\right)  },$ & if
$x\in\Omega\left(  t\right)  ,$\\
$0,$ & otherwise.
\end{tabular}
\ \ \ \ \ \right.  \label{eq:heat-out-spread}%
\end{equation}
The coefficients $R_{0}$, $S_{\max}$, $a$, $b$, $W$, and $A$, which
characterize the fuel, are determined from laboratory experiments.

This model is developed in \cite{Clark-2004-DCA}, where a tracer scheme is
used to advance the fireline. Tracer schemes do well when advecting the shape
of the fire in a wind, but require complicated code for numerous special cases
when the topology of the fireline changes, such as when fire fronts merge.
Also, tracer schemes are not well suited for data assimilation. Changing the
location of a fireline represented by tracers requires a special, complicated
code, and the state of the model cannot be adjusted by making corrections to
gridded arrays, as is usual in data assimilation methods.

Hence, another implementation of the propagation of the fireline by the level
set method is considered. The level set method evolves a function $\psi
=\psi\left(  x,t\right)  $, called the \emph{level set function}, such that
the burning area is $\Omega\left(  t\right)  =\left\{  x:\psi\left(
x,t\right)  <0\right\}  $ and the fireline is the level set
\[
\Gamma(t)=\left\{  x:\psi\left(  x,t\right)  =0\right\}  .
\]
The level set function satisfies the differential equation%
\begin{equation}
\frac{\partial\psi}{\partial t}+S\left(  x\right)  \left\Vert \nabla
\psi\right\Vert =0, \label{eq:level}%
\end{equation}
which is solved numerically. The state of the model consists of the level set
function $\psi$ and the ignition time $t_{i}$, given as their values on grid
nodes. The remaining fuel is determined from the ignition time by
(\ref{eq:burning}), but it could be alternatively maintained as a separate
variable instead of the time of ignition, which might have some advantages in
data assimilation.

For further details, see \textquotedblleft Level set-based wildland fire
model.\textquotedblright\ Since all quantities are mesh arrays and additive
changes to their values are meaningful, the state can be modified by data
assimilation methods in the usual manner.\ 

\subsection{Coupling fire and weather models}

The fire model can be either a physical formulation such as (\ref{eq:heat}) --
(\ref{eq:reaction}) or a semi-empirical formulation like (\ref{eq:Rothermel})
-- (\ref{eq:level}). So far, we have implemented the coupling with the
semi-empirical model only. In either case, the fire model takes as input the
horizontal wind velocity $\overrightarrow{v}$ and it outputs the heat flux
$H$, given by (\ref{eq:heat-out-pde}) or (\ref{eq:heat-out-spread}). The
Rothermel formula (\ref{eq:Rothermel}) was originally developed from
laboratory measurements assuming undisturbed external wind near the fire. In a
two-way coupled model, undisturbed winds are not available due to the feedback
of the fire onto the atmosphere. Instead, the horizontal wind velocity at a
small distance behind the fireline should be used. The fire mesh is generally
finer than the atmospheric mesh, so the wind is interpolated to the nodes of
the fire mesh, and the heat flux is aggregated over the cells of the fire mesh
that make up one cell of the atmospheric mesh.

In \cite{Clark-2004-DCA}, the Clark-Hall atmospheric code
\cite{Clark-1996-DSC} is coupled with the semi-empirical model
(\ref{eq:Rothermel}) -- (\ref{eq:level}), implemented by a tracer scheme. This
code can nest and refine meshes from synoptic scale ($10^{2}$ km) to fire
scale ($10$ m) horizontally and also refine meshes vertically; however, the
tracer scheme is not suitable for data assimilation, and export and import of
the state, needed for data assimilation, is not supported. In contrast, the
Weather Research and Forecasting code (WRF, \cite{WRF}) coupled with the
semi-empirical fire model implemented by the level set method is used in the
current work. WRF is a standard for weather forecasting, it supports
distributed memory execution, and provides import and export of the state for
data assimilation.

For the physical fire model (\ref{eq:heat}) -- (\ref{eq:reaction}), the state
is the temperature $T$ and the fuel supply $F$. The state of the
semi-empirical fire model (\ref{eq:Rothermel}) -- (\ref{eq:level}) is the
level set function $\psi$ and the ignition time $t_{i}$, which records the
level set function history. At the beginning of an atmospheric time step, the
wind is interpolated from the atmospheric mesh to the nodes of the fire mesh.
The fire model is then advanced one or more internal time steps to the end of
the atmospheric time step. The maximum time step in the fire model is limited
by the stability restriction of the numerical scheme. In the computations
reported here, the time step for the atmospheric model was short enough for
the fire model, so only one time step of the fire model was done. After
advancing the fire model, the total heat flux $H$ generated over the
atmospheric time step is inserted in the atmospheric model. The heat flux is
split into sensible heat flux (a transfer of heat between the surface and air
due to the difference in temperature between them) and latent heat flux (the
transfer of heat due to the phase change of water between liquid and gas) in
the proportion given by the fuel type and its moisture. A difficulty is that
it is not possible to apply a heat flux directly as a boundary condition to an
atmospheric model on the derivatives of the corresponding physical field (air
temperature or water vapor contents) because the atmospheric codes used here
do not support flux boundary conditions. The atmospheric codes use numerical
methods explicit in time, and imposing a flux boundary condition with an
explicit timestepping method directly would add all heat or vapor influx into
a boundary layer one cell thick, requiring diffusion and convection in later
time steps to transport the heat further into the domain. Thus, the flux would
result in a large non-physical change in temperature or vapor concentration in
the boundary layer, and its effect would be resolved only over a number of
time steps, progressing only one mesh cell away from the boundary in every
time step. Implicit timestepping methods do not suffer from such ill effect,
because the global problem solved in each time step forces all physical fields
to be in balance at the end of the time step. Therefore, an empirical
procedure is used, and the flux is inserted by modifying the temperature and
water vapor concentration over a depth of many cells, with exponential decay
away from the boundary. This decay mimics the distribution of temperature and
water vapor fields arising from the vertical flux divergence, which is
supported by infrared observations of the dynamics of crown fires in
\cite{Coen-2004-IIC}.

A typical horizontal mesh size for WRF for weather modeling is about $1$ km,
based on the spatial scales needed for modeling clouds, although the numerical
approximations of the dynamical equations support finer resolution. WRF
supports refinement by nested meshes. The finest atmospheric mesh interfaces
with the fire. However, meshes can be nested only horizontally, so even the
finest mesh still goes from the surface of the Earth all the way to the top of
the model domain, which is an unnecessary computational expense.

The recommended time step in WRF is proportional to the finest atmospheric
mesh step, and is $1$ s for the mesh step of $1$ km. However, the need for
fire resolution demands a finer fire mesh, and a correspondingly finer
atmospheric mesh to capture properly the feedback of the winds on the shape of
the fire. In the computational experiments reported in this paper, we have
used fire mesh step size $6$ m on a $420$ by $420$ grid, and horizontal the
atmospheric mesh step size $60$ m, on a $42$ by $42$ atmospheric mesh, for a
$2,400$ m by $2,400$ m physical domain. The atmospheric mesh had $41$
horizontal layers from the Earth surface to the top at $1,500$ m, graded so
that the thickness of the lowest atmospheric mesh was $25$ m. The time step of
the atmospheric model was dictated by its stability restrictions; however,
$250$ ms was observed to be sufficient rather than the $60$ ms that would be
recommended. The dimension of the state vector, including both the fire and
the atmosphere states, was $875,404$. The complete state, as exported by WRF,
contains many constant arrays, and is about $20$ times bigger. A more
realistic fire model may easily require a domain $30$ km by $30$ km, since
fires have been known to spread that far. Still with $41$ horizontal layers,
such problem would have about $525$ by $525$ nodes on the atmospheric mesh and
$5250$ by $5250$ nodes on the fire mesh. The dimension of the state vector
would be about $1.4\cdot10^{8}$. The code runs in single precision with $4$
bytes per floating point number, giving the size of the state vector about
$500$ MB and the complete WRF state about $10$ GB.

WRF is usually run as a hierarchy of coupled models on nested grids to bridge
the scales between a global atmospheric forecast and the microscale model in
the area of interest. A nested model that would incorporate mesoscale weather
effects of the fire would require an additional, coarser grid, over a larger
area, approximately doubling the state size.

%The large difference in the scales of the weather model and the fire model is
%definitely a concern, as it may affect the quality of the modeling of the
%feedback of fire on the atmosphere. Therefore, an intermediate Computational
%Fluid Dynamics (CFD) code between the fire scale and atmospheric scale is
%considered for future work. The WRF dynamic core, which is a CFD solver
%without any physics parametrization, can be used for that purpose similarly
%as in \cite{Cunningham-2005-CVS}.

\section{High-performance implementation of the EnKF}

In this section, we describe the base EnKF method before our modifications
specific to the fire application, and its parallel implementation. We use the
EnKF method from \cite{Burgers-1998-ASE}, which involves randomized data and
full data error covariance. Since the number of degrees of freedom in the
model state as well as the number of data points is large and the code needs
to run much faster than real time, efficient massively parallel implementation
is essential. This section is based on \cite{Mandel-2006-EIE}, where more
details can be found. We start by recalling the Kalman filter formulas.

\subsection{The Kalman formulas as Bayesian update for Gaussian distributions}

Let $\mathbf{x}$ denote the $n$-dimensional state vector of a model, and
assume that it has Gaussian probability density function (pdf) $p(\mathbf{x})$
with mean $\mathbf{\mu}$ and covariance $Q$. This probability distribution,
called the \emph{prior}, was evolved in time by running the model and now is
to be updated to account for new data. Given state $\mathbf{x}$, the data
$\mathbf{d}$ are assumed to have Gaussian distribution $p\left(
\mathbf{d}|\mathbf{x}\right)  $, called \emph{data likelihood}, with
covariance $R$ and mean $H\mathbf{x}$. The matrix $H$ is called an
\emph{observation matrix}. The value $H\mathbf{x}$ is what the value of the
data would be for the state $\mathbf{x}$ in the absence of data errors, and
the data covariance describes an estimate of the data errors; that is, the
state should satisfy (in statistical sense) the \emph{observation equation}
\begin{equation}
H\mathbf{x}=\mathbf{d}. \label{eq:obs-eq}%
\end{equation}
The pdf of the state and the data likelihood are combined to give the new
probability density of the system state $\mathbf{x}$ conditional on the value
of the data $\mathbf{d}$ (the \emph{posterior}) by the Bayes theorem%
\[
p\left(  \mathbf{x}|\mathbf{d}\right)  =\frac{p\left(  \mathbf{y}%
|\mathbf{d}\right)  p(\mathbf{y})}{\int p\left(  \mathbf{z}|\mathbf{d}\right)
p(\mathbf{z})d\mathbf{z}}.
\]
The data $\mathbf{d}$ is fixed once it is received, so denote the posterior
state by $\mathbf{\hat{x}}$ instead of $\mathbf{x}|\mathbf{d}$ and the
posterior pdf by $p\left(  \mathbf{\hat{x}}\right)  $. By algebraic
manipulations, one can show \cite{Anderson-1979-OF} that the posterior pdf is
also Gaussian with the posterior mean $\mathbf{\hat{\mu}}$ and covariance
$\hat{Q}$ given by the Kalman update formulas%
\[
\mathbf{\hat{\mu}}=\mathbf{\mu}+K\left(  \mathbf{d}-H\mathbf{\mu}\right)
,\quad\hat{Q}=\left(  I-KH\right)  Q,
\]
where%
\[
K=QH^{\mathrm{T}}\left(  HQH^{\mathrm{T}}+R\right)  ^{-1}%
\]
is called the Kalman gain matrix.

\subsection{The EnKF as a Monte Carlo approximation}

Suppose $\mathbf{x}_{1},\ldots,\mathbf{x}_{N}$ is a random sample from the
prior and the matrix $X$ consists of $\left[  \mathbf{x}_{1},\ldots
,\mathbf{x}_{N}\right]  $ as columns, $X=\left[  \mathbf{x}_{1},\ldots
,\mathbf{x}_{N}\right]  =\left[  \mathbf{x}_{i}\right]  $. Replicate the data
$\mathbf{d}$ into an $m\times N$ matrix $D=\left[  \mathbf{d}_{1}%
,\ldots,\mathbf{d}_{N}\right]  =\left[  \mathbf{d}_{i}\right]  $ so that each
column $\mathbf{d}_{i}$ consists of the data vector $\mathbf{d}$ plus a random
vector from the $n$-dimensional normal distribution with zero mean and
covariance matrix $R$. Then the columns of
\begin{equation}
\hat{X}=X+K(D-HX) \label{eq:enkf-exact}%
\end{equation}
form a random sample from the posterior distribution. The EnKF consists of the
Bayesian step (\ref{eq:enkf-exact}) with the following approximations. First,
the mean and the covariance of the pdf of the prior are replaced by the sample
mean and covariance computed from the ensemble. Furthermore, the ensemble
members are really not a random sample, because they are not independent --
the EnKF ties them together in every Bayesian step. With these approximations,
the EnKF posterior ensemble is%
\begin{equation}
X^{a}=X+CH^{T}\left.  \underbrace{\left(  HCH^{T}+R\right)  }_{P}\right.
^{-1}(D-HX), \label{eq:enkf-update}%
\end{equation}
where%
\begin{equation}
C=\frac{AA^{T}}{N-1},\quad A=X-\frac{1}{N}\left(  X\mathbf{e}_{N\times
1}\right)  \mathbf{e}_{1\times N}, \label{eq:enkf-mat}%
\end{equation}
and $\mathbf{e}$ denotes the matrix of all ones of the indicated size. From
(\ref{eq:enkf-mat}), it follows that $C=XM$ for some matrix $M$. Consequently,
the posterior ensemble consists of linear combinations of the prior ensemble.

\subsection{Efficient parallel implementation of EnKF formulas}

Note that since $R$ is a covariance matrix, it is always positive
semidefinite, and typically positive definite and well conditioned. Then
$P^{-1}$ exists and (\ref{eq:enkf-update}) can be implemented efficiently by
the Choleski decomposition. Equations (\ref{eq:enkf-update}) --
(\ref{eq:enkf-mat}) are the same as in~\cite{Burgers-1998-ASE}. There are
versions of EnKF \cite{Evensen-2007-DAE} which use the sample covariance of
the perturbed data, and the inverse then needs to be replaced by a
pseudoinverse, computed by the Singular Values Decomposition (SVD). SVD is an
iterative algorithm and much more expensive than a direct method involving
only matrix multiplication and Choleski decomposition. However, additional
efficiency in SVD-based methods may be gained by using a reduced rank
pseudoinverse, i.e., dropping small singular values \cite{Evensen-2007-DAE}.
In contrast, the implementation discussed here consists of straightforward
linear algebra, it does not change the mathematical algorithm from
\cite{Burgers-1998-ASE}, and it does not involve any tolerances that might
need to be tuned by the user.

For a large number $m$ of data points, such as in the assimilation of gridded
data or images, the Choleski decomposition of the matrix $P$ becomes a
bottleneck, as it requires $O\left(  m^{3}\right)  $ operations. A better way
is possible when the data error covariance matrix $R$ is diagonal (which is
the case when the data errors are uncorrelated an doften satisfied in
practice), or at least cheap to decompose (such as block diagonal or banded
due to limited covariance distance). Using the Sherman-\allowbreak
Morrison-\allowbreak Wood\-bury formula~\cite{Hager-1989-UIM}
\[
(R+UV^{T})^{-1}=R^{-1}-R^{-1}U(I+V^{T}R^{-1}U)^{-1}V^{T}R^{-1}%
\]
with $U=\frac{1}{N-1}HA$ and $V=HA$ gives%
\begin{align}
P^{-1}  &  =\left(  R+\frac{1}{N-1}HA\left(  HA\right)  ^{T}\right)
^{-1}\nonumber\\
&  =R^{-1}\left[  I-\frac{1}{N-1}\left(  HA\right)  \left(  I+\left(
HA\right)  ^{T}R^{-1}\frac{1}{N-1}\left(  HA\right)  \right)  ^{-1}\left(
HA\right)  ^{T}R^{-1}\right]  , \label{eq:P-inv}%
\end{align}
which requires only the solutions of systems with the matrix $R$ (assumed to
be cheap) and of a system of size $N$ with $m$ right-hand sides. When $R$ is
diagonal, the evaluation of (\ref{eq:enkf-update}) with (\ref{eq:P-inv}) costs
$O\left(  N^{3}+mN^{2}+nN^{2}\right)  $ operations, which is suitable both for
a large number $n$ of the degrees of freedom of the state and a large number
$m$ of data points. Instead of computing the inverse of a matrix and
multiplying by it, it is much better (several times cheaper, more accurate,
and standard practice) to compute the Choleski decomposition of the matrix and
implement the multiplication by the inverse as solution of a linear system
with many simultaneous right-hand sides~\cite{Golub-1989-MAC}.

When $R$ is diagonal, it is also possible to assimilate the data points
sequentially, since then the observations are independent, but level 3
operations (matrix-matrix, that is, many vectors at a time) are more efficient
than repeated level 2 matrix-vector operations on one vector at a time.

Since the EnKF formulas are written as matrix algebra with dominant Level 3
(dense matrix-matrix) operations~\cite{Golub-1989-MAC}, they are suitable for
efficient implementation using software packages such as LAPACK
\cite{Anderson-1999-LUG} on serial and shared memory computers\ and ScaLAPACK
\cite{Dongarra-1996-ST} on distributed memory computers. Such implementations
have the advantage that they keep the high-performance computing and
methodological issues separate. In particular, they do not change the
mathematical method, unlike, e.g., \cite{Keppenne-2002-ITM}, where parallelism
is achieved by modifying the EnKF based on the spacial locality of the unknowns.

Finally, it is inconvenient to construct and operate with the matrix $H$
explicitly; instead, we wish to evaluate a function $h(x)$ of the form
\begin{equation}
h(\mathbf{x})=H\mathbf{x+f}, \label{eq:obs-function}%
\end{equation}
where $H$ and $\mathbf{f}$ are fixed but unknown, and instead of
(\ref{eq:obs-eq}), use the observation equation $h(\mathbf{x})=\mathbf{d}%
^{\prime}$, where $\mathbf{d=}$ $\mathbf{d}^{\prime}-\mathbf{f}$. The function
$h$ is called the \emph{observation function} or, in the inverse problems
context, the \emph{forward operator}. The value of $h(\mathbf{x})$ is what the
value of the data $\mathbf{d}^{\prime}$ would be for the state $\mathbf{x}$
assuming the measurement is exact. It is easy to see that $H$ occurs in
(\ref{eq:enkf-update}), (\ref{eq:enkf-mat}), and (\ref{eq:P-inv}) only in the
expressions $HA$ and $HX-D$, which is obtained by random perturbations of the
columns $H\mathbf{x}_{i}-\mathbf{d}$. But the columns of the matrix-matrix
product $HA$ and the residuals $H\mathbf{x}_{i}-\mathbf{d}$ can be evaluated
by calling $h$ once on every ensemble member%
\begin{align*}
\left[  HA\right]  _{i}  &  =H\mathbf{x}_{i}-H\frac{1}{N}\sum_{j=1}%
^{N}\mathbf{x}_{j}=h\left(  \mathbf{x}_{i}\right)  -\frac{1}{N}\sum_{j=1}%
^{N}h\left(  \mathbf{x}_{j}\right)  ,\\
H\mathbf{x}_{i}-\mathbf{d}  &  =H\mathbf{x}_{i}-\left(  \mathbf{d}^{\prime
}-\mathbf{f}\right)  =h\left(  \mathbf{x}_{i}\right)  -\mathbf{d}^{\prime}.
\end{align*}

Such approach is commonly used for a nonlinear observation function, such as
the position of a hurricane vortex~\cite{Chen-2007-AVP}. Essentially, this
approach approximates the observation function by the linear interpolation
from its values at the ensemble members.

\subsection{EnKF in a high-performance computing environment}

Our parallel implementation uses a distributed memory parallel linear algebra
software layer built on top of ScaLAPACK. The ensemble is intepreted by the
linear algebra software as a distributed matrix. Each column is the state
vector of one ensemble member, and each process has one or more columns in its
memory. ScaLAPACK then operates on the ensemble like on any other distributed
matrix. The distribution of the intermediate matrices in the computations is
determined by the requirements of ScaLAPACK, and it is not limited to column
distribution. Correct distributions must be used for parallel scalability of
the linear algebra algorithms, i.e., to speed up the computations optimally
with the number of processors. For example, parallel scalability of Choleski
decomposition in ScaLAPACK requires a decomposition into approximately square tiles.

\begin{figure}[ptb]
\begin{center}
\includegraphics[width=6in]{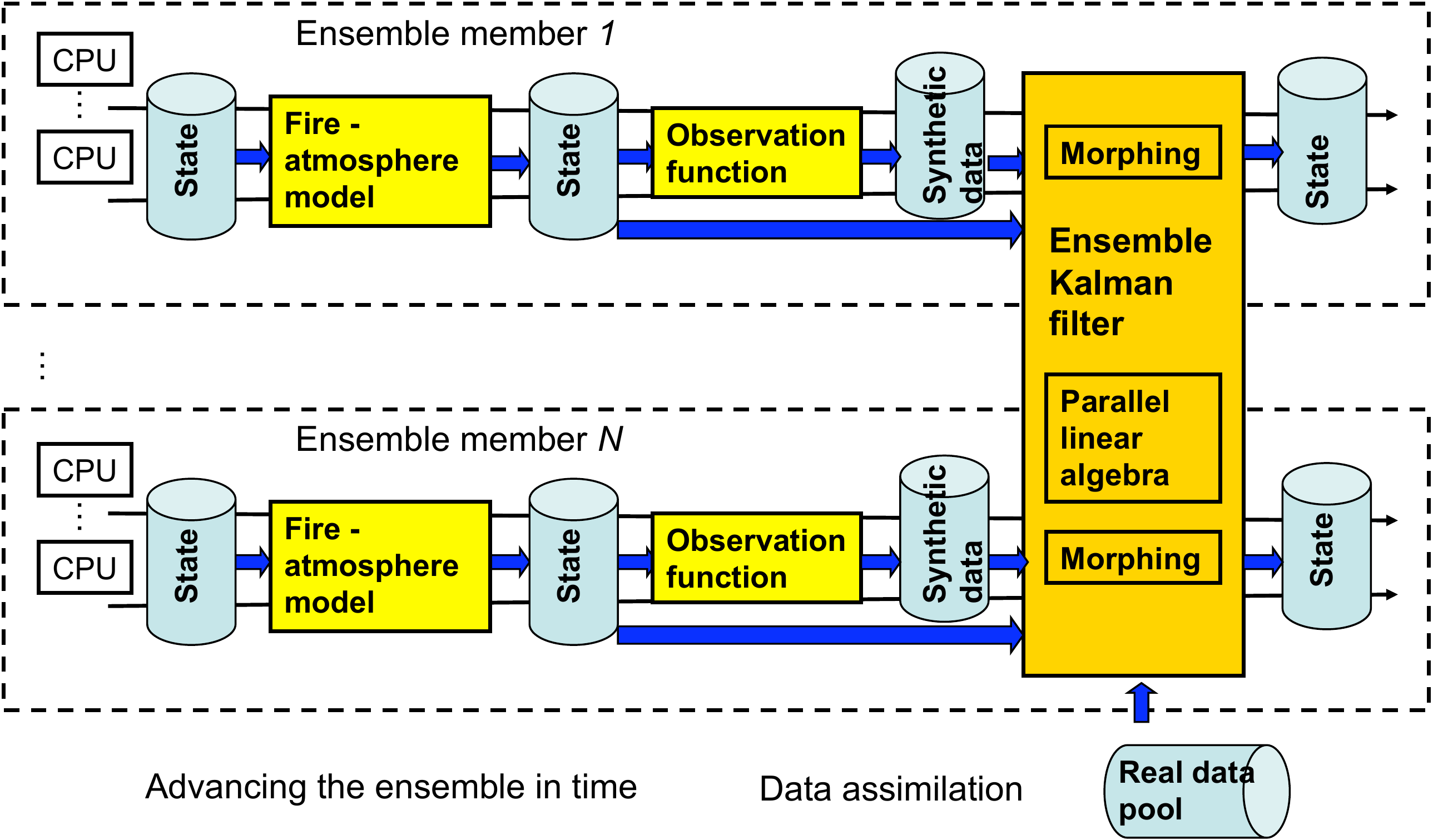} \newline
\end{center}
\caption{Parallel ensemble management scheme. In each cycle, all ensemble
members are advanced in time, each member by a separate run of the simulation
code. The simulation code is a parallel application, which executes on a
collection of processors, both for speed and distribution of the memory use.
When enough processors are available, some or all ensemble members are
advanced concurrently. After all ensemble members are advanced in time, the
value of the observation function is evaluated for each ensemble member and
the EnKF code is started on all available nodes. The EnKF code loads the state
of the ensemble members and the values of the observation function from the
files, reads the data to be assimilated from data files, and performs the
analysis calculations using high-performance parallel linear algebra. The
modified states of ensemble members are written to files and the cycle
repeats.}%
\label{fig:parallel}%
\end{figure}

The analysis cycles are run in a shell script loop. Each ensemble member state
is maintained in a separate files. The model, the EnKF code, and the
evaluation of the observation function are in separate executables, which
communicate by files. Each ensemble member runs on a separate group of
processors, and the EnKF running on all processors ties them
together~(Fig.~\ref{fig:parallel}).

Running the EnKF and the simulations as a single parallel application with
data passed in memory only is also possible, as done \cite{Mandel-2007-DDD}
for the standalone reaction-diffusion equation fire model. However, running
the EnKF and the simulations separately affords more flexibility, does not
force integration of the codes into a single executable, and allows for
natural checkpointing and restart of the data assimilation scheme.

For efficiency, on computer clusters with local filesystems on nodes, the
files with the state of a member should be kept in local files on the nodes
used to advance that member. Some computer systems, such as IBM BlueGene, have
a single high-performance file system, designed for massively parallel access,
with multiple independent high-speed links from subsets of nodes. In that
case, simulations running on separate groups of nodes can access their files
on the common filesystem without interference with each other.

\section{Assimilation of data into fire models}

Wildfire data is often sparse in space and time and innacuracies in the models
and the initial conditions are inevitable. So, the model state can differ
significantly from the data that is being assimilated. Yet, the simulation
must recover and continue. Thus, an important quality of data assimilation
methods is their ability to make large corrections to the state without
breaking the assimilation method or the model.

The EnKF applied to either fire model described above fails quickly
\cite{Mandel-2008-WFM,Beezley-2008-MEK}. One reason is that errors in fire
simulations often involve an inaccurate location of the fire. The task of the
EnKF to find the best fit to the data is then futile, because the EnKF makes
amplitude corrections rather than position corrections.

Another aspect of the failure of the EnKF that the EnKF is based on the
assumption that all probability distributions involved are at least
approximately Gaussian. While the location of the fire may have an error
distribution that is approximately Gaussian, this phenomenon is certainly not
the case for the value of the state (such as the temperature) at a~given
point. Instead, the probability distribution of the state at a given point
near the fireline is concentrated around the burning and the not burning states.

Finally, fire models are irreversible and non dissipative, and, consequently,
they behave quite differently from the usual atmospheric and oceanic models
that EnKF is applied to. The mode of failure of the EnKF for both models is
similar. Statistical variability of the EnKF corrections causes spurious
fires, which continue burning and may even grow instead of conveniently going
away like, e.g., spurious plumes in pollution models, which dissipate readily.
In physics based models, such as the reaction-diffusion model described above,
the EnKF corrections result in nonphysical states, which cause the model to
break down.

\subsection{Regularized EnKF}

\begin{figure}[ptb]
\begin{center}
\includegraphics[width=5in]{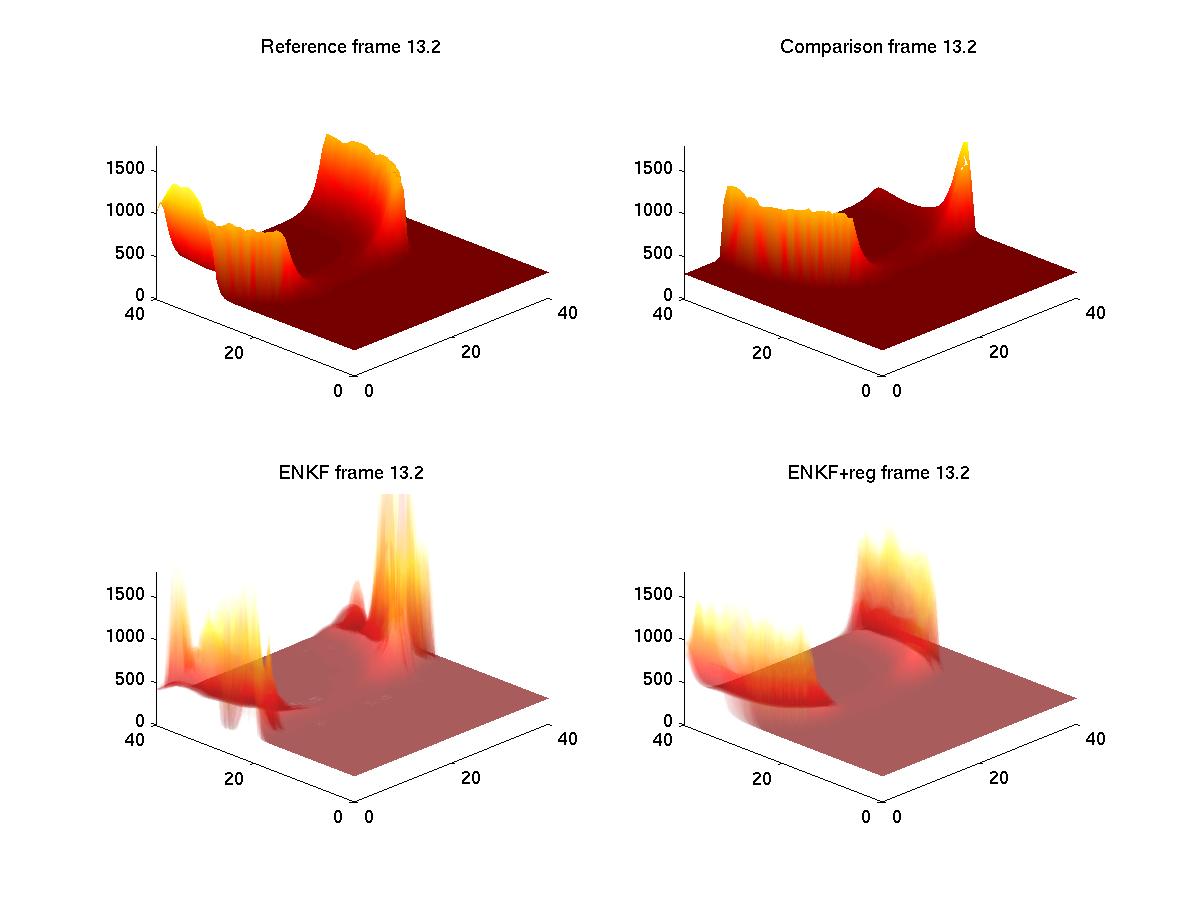} \newline
\end{center}
\caption{Example of the effect of regularization by penalization of large
change in gradients by the Bayesian update on propagation of fire. Circular
spreading fire hits a fuel break in the middle of the domain. The horizontal
axes are distance in m. The vertical axis is temperature in K. False color is
generated from the temperature with shading for depth perception. The
reference solution is the simulated data. In the comparison solution, the fire
was started from intentionally incorrect location. The ENKF panel is the
result of data assimilation after 13 cycles. The ENKF+reg panel is the
ensemble after data assimilation after 13 cycles with regularization imposed
in each cycle. The ensembles have 100 members and they are visualized as
superposition of transparent images of their members. The observation function
was the temperature field on the whole grid. Both ensembles succeed in moving
toward the data, but the regularized ensemble avoids the non-physical
artefacts, such as extremely small and large temperatures.}%
\label{fig:reg}%
\end{figure}

The nonphysical states from the EnKF analysis step in the physics based model
are characterized by the values of the physical fields out of bounds, such as
the spikes in the temperature field in the lower left panel in Fig.
\ref{fig:reg}. Therefore, it is natural to consider penalizing the large
nonphysical values, hoping to stabilize the computation. This approach was
tried, but without much success, because the nonphysical values are not all
that different from valid ones. However, the spatial gradients in the spikes
in the state from EnKF are much larger than the gradients normally seen in a
physically reasonable solution. Thus, the regularized
EnKF~\cite{Johns-2008-CEK} was developed which penalizes large changes in
spatial gradients in the Bayesian update. One possible implementation is to
add a penalty term involving the norm of the spatial gradients of the solution
to the objective function in the minimization formulation of EnKF, much like
Tikhonov regularization for ill-posed problems. However, the regularization
can be implemented simply by adding an artificial observation
$\bigtriangledown u=\bigtriangledown\bar{u}$ to EnKF, where $u$ is one of the
fields in model state (such as the temperature), $\bigtriangledown$ is the
spatial gradient, computed by finite differences, and $\bar{u}$ is the same
field in the mean of the forecast ensemble. The amount of penalty is
controlled by the error covariance for the observation; smaller error
covariance means stronger penalization of large changes in the gradient.
Because the added observation is independent of other data in the Bayesian
update, it is easily implemented by running the EnKF\ formulas twice, thus
requiring very little new code.

The regularization technique does have a stabilizing effect on the simulations
in the ensemble (Fig. \ref{fig:reg}), but finding the proper value of the
penalty constant is tricky and the regularization does not improve much the
ability of the EnKF to track the data. The posterior ensemble is made out of
linear combinations of the prior ensemble, and if a reasonably close location
and shape of the fire cannot be found between the linear combinations, the
data assimilation is simply out of luck, and the ensemble cannot approach the
data. From that point on, the ensemble evolves essentially without regard to
the data.

\subsection{Morphing EnKF}

\begin{figure}[ptb]
\begin{center}%
\begin{tabular}
[c]{cc}%
\includegraphics[width=2.6in]{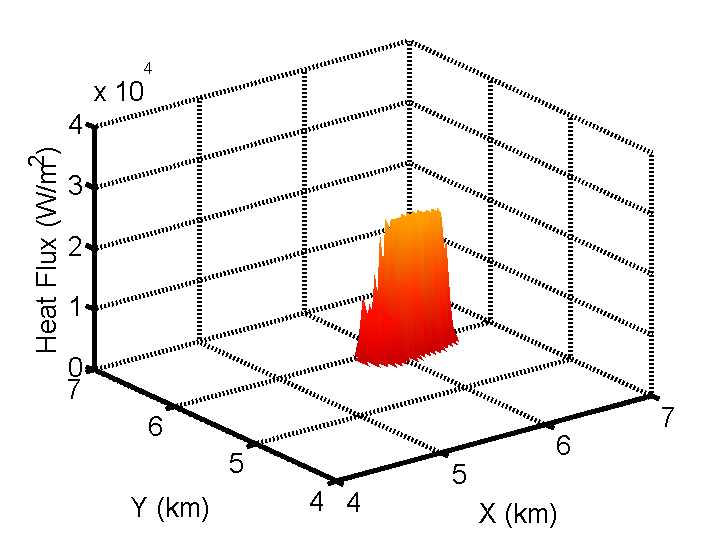} &
\includegraphics[width=2.6in]{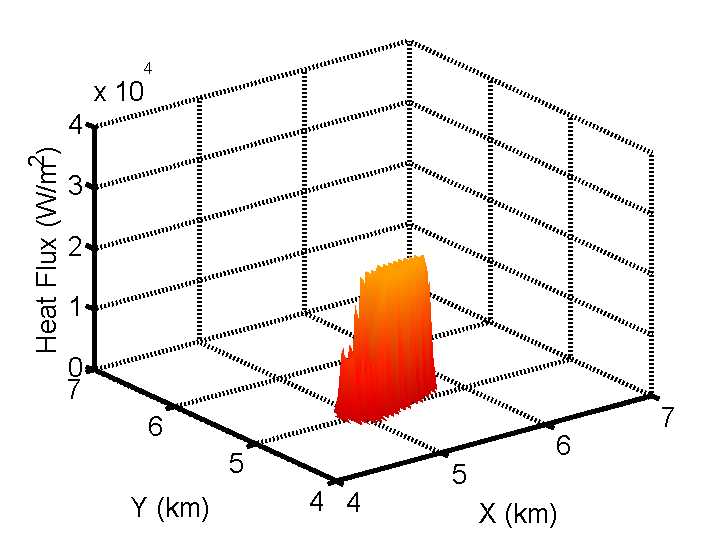}\\
(a) & (b)\\
\includegraphics[width=2.6in]{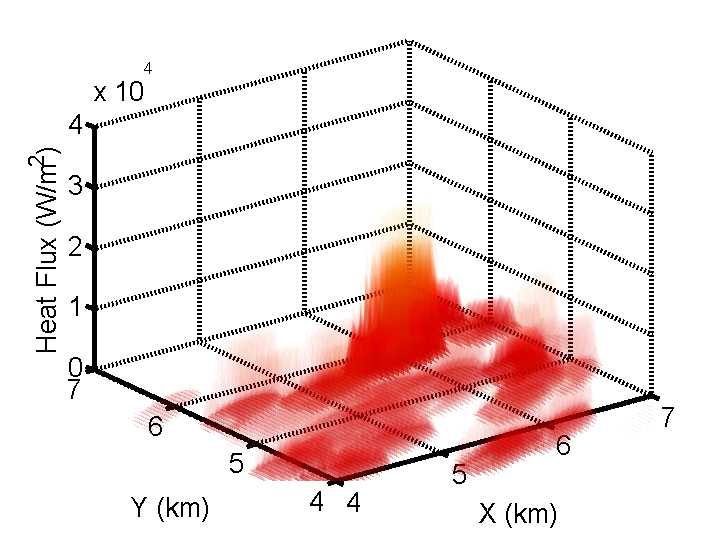} &
\includegraphics[width=2.6in]{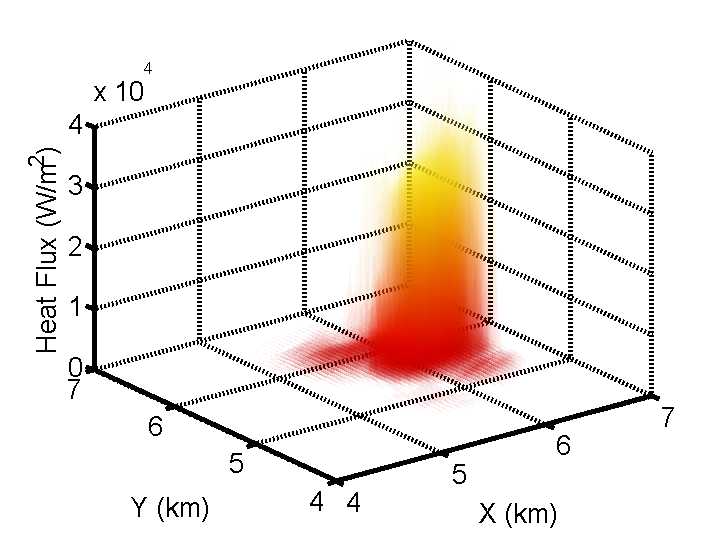}\\
(c) & (d)
\end{tabular}
\newline
\end{center}
\caption{The morphing EnKF applied to the fireline propagation model. False
color is the output heat flux with shading for depth perception. The reference
solution (a) is the simulated data. The initial ensemble was created by a
random perturbation of the comparison solution (b), where the fire was ignited
at an intentionally incorrect location. The standard ENKF panel (c) is the
result of data assimilation of the time from ignition after running the model
for 1000 seconds. The morphing EnKF panel (d) is the result with image
registration determined from the time from ignition and applied to all of the
model variables. The ensembles have 25 members each and they are visualized as
superposition of transparent images of heat fluxes of their members. The
registration was done on the atmospheric grid with the fire heat flux as the
observation function, but the atmospheric model was not used. The standard
EnKF ensembles diverges from the data, while the morphing EnKF ensemble keeps
closer to the data.}%
\label{fig:morph-level}%
\end{figure}

A level set fire spread model makes it possible to manipulate the location of
the fire easily by changing the level function just like any other physical
field. Yet, the hope that the level set representation of the fire region
would be all that would be needed for a succesful fire data assimilation did
not pan out: the EnKF causes a number of spurious fires even with the level
set model, as seen in the lower left panel in Fig. \ref{fig:morph-level}.

There is clearly a~need to adjust the simulation state by a change of position
explicitly rather than by an additive correction. We use techniques from image
processing for this purpose. Moving and stretching one given image to become
another given image is known as registration \cite{Brown-1992-SIR}. Classical
registration methods require a user to pick which points are to be transformed
into which points, but fully automatic methods now exist. Once the two images
are registered, one can easily create intermediate images, which is known as
morphing. The intermediate images can be created in such a way that they can
be used instead of linear combinations of states in EnKF. The resulting method
\cite{Beezley-2008-MEK} provides both additive and position correction
combined in a natural manner. Related methods in the literature include
transformation of the space by a~low-order polynomial mapping
\cite{Alexander-1998-UDW} and solving a differential equation to find an
alignment mapping as a preprocessing to an additive correction
\cite{Lawson-2005-AEM,Ravela-2007-DAF}.

The basic ingredient in the morphing EnKF is automatic registration, borrowed
from image processing. Consider two functions $u_{0}$, $u_{1}$, representing
the same physical field, such as the temperature, or the level set function,
from two states of a wildfire model. The functions are represented by gridded
arrays on the fire problem domain, and considered interpolated away from grid
nodes. Let $x=\left(  x_{1},x_{2}\right)  $ be coordinates of a point in the
fire problem domain on the surface of the Earth. The registration problem is
to find a mapping $T$ $:x\mapsto T\left(  x\right)  $ such that the change of
variable $x\mapsto x+T\left(  x\right)  $ transforms $u_{0}$ into a function
approximately equal to $u_{1}$, $u_{0}\left(  x+T\left(  x\right)  \right)  $
$\approx u_{1}\left(  x\right)  $. This process can be written in the compact
form
\begin{equation}
u_{1}\approx u_{0}\circ\left(  I+T\right)  \text{ on }D, \label{eq:reg}%
\end{equation}
where $\circ$ denotes the composition of mappings, and $I:x\mapsto x$ is the
identity mapping. The mapping $I+T$ is called the \emph{registration mapping},
and the mapping $T$ is called \emph{warping}. The reason for writing the
registration mapping as $I+T$ is that the zero warping $T=0$ is the neutral
element of the operation of addition, and so linear combinations of warpings
have a~meaningful interpretation as blends of the warpings.

To avoid unnecessarily large warping, $T$ must be as close to zero and as
smooth as possible%
\begin{equation}
T\approx0,\quad\nabla T\approx0, \label{eq:small-T}%
\end{equation}
where $\nabla T$ denotes the matrix of the first derivatives (the Jacobian
matrix) of $T$. In addition, we require that the registration mapping $I+T$ is
one-to-one, so that the inverse $\left(  I+T\right)  ^{-1}$ exists. For a
fully automatic method to construct such a mapping $T$, see \textquotedblleft
Image registration.\textquotedblright\ The mapping $T$ is represented by its
values on the grid and interpolated away from the grid points.

Once the registration mapping $I+T$ is found, one can construct intermediate
functions $u_{\lambda}$ between $u_{0}$ and $u_{1}$ by the morphing formula
\cite{Beezley-2008-MEK}
\begin{equation}
u_{\lambda}=\left(  u+\lambda r\right)  \circ\left(  I+\lambda T\right)
,\quad0\leq\lambda\leq1, \label{eq:ulambda}%
\end{equation}
where
\begin{equation}
r=u_{1}\circ\left(  I+T\right)  ^{-1}-u_{0}. \label{eq:inverse_res}%
\end{equation}
The original functions $u_{0}$ and $u_{1}$ are recovered by choosing in
(\ref{eq:ulambda}) $\lambda=0$ and $\lambda=1$. The formula (\ref{eq:ulambda})
blends naturally the change in position and value. It is more complicated and
expensive than alternatives, such as $u_{\lambda}=u_{0}\circ\left(  I+\lambda
T\right)  $, which modifies the position only, or $u_{\lambda}=u_{0}%
\circ\left(  I+\lambda T\right)  +\lambda\left(  u_{1}-u_{0}\circ\left(
I+T\right)  \right)  $, which combines value and position correction but the
value correction is always in a fixed place. However, the results obtained
with (\ref{eq:inverse_res}) are much more satisfactory, and the inverse
mapping can be evaluated by a fast inverse interpolation: if $y_{i}$ are grid
nodes and $x_{i}=\left(  I+T\right)  y_{i}$, then $y=\left(  I+T\right)
^{-1}x$ is obtained by interpolating the values of $y_{i}$ on the nonuniform
grid formed by $x_{i}$.

The morphing EnKF works by transforming the member states into extended states
$\left[  r,T\right]  $ consisting of additive and position components that
encode the difference of the member from some base state $u_{0}$. The EnKF
formulas are then run on the extended states $\left[  r,T\right]  $.
Afterwards, the extended states are converted back by (\ref{eq:ulambda}) and
advanced in time. The base state $u_{0}$ can be created by taking the average
of the extended states $\left[  r,T\right]  $ after the EnKF update,
transforming back, and advancing in time. Alternatively, the base state can be
chosen as one of the ensemble members, or, for short simulation times, the
base state may be fixed as the initial state that the ensemble was created from.

\begin{figure}[ptb]
\begin{center}%
\begin{tabular}
[c]{cc}%
\includegraphics[width=2.6in]{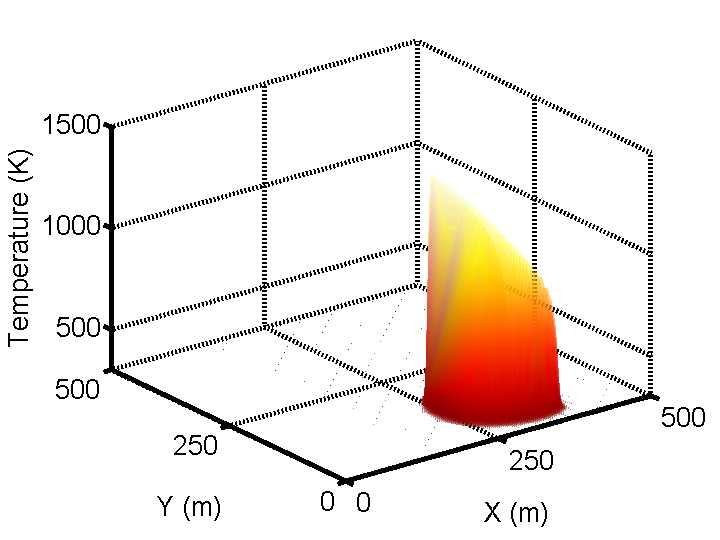} &
\includegraphics[width=2.6in]{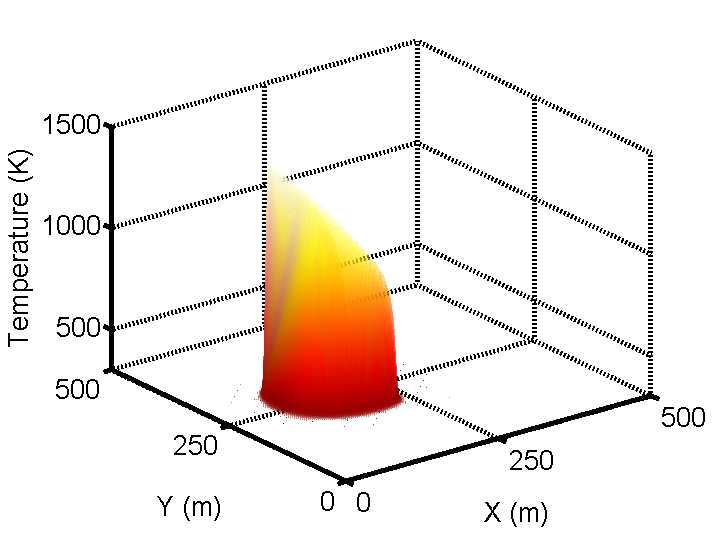}\\
(a) & (b)\\
\includegraphics[width=2.6in]{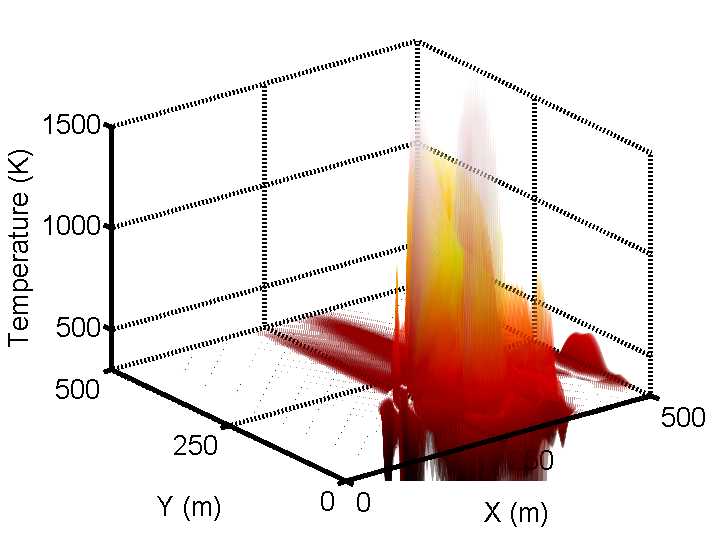} &
\includegraphics[width=2.6in]{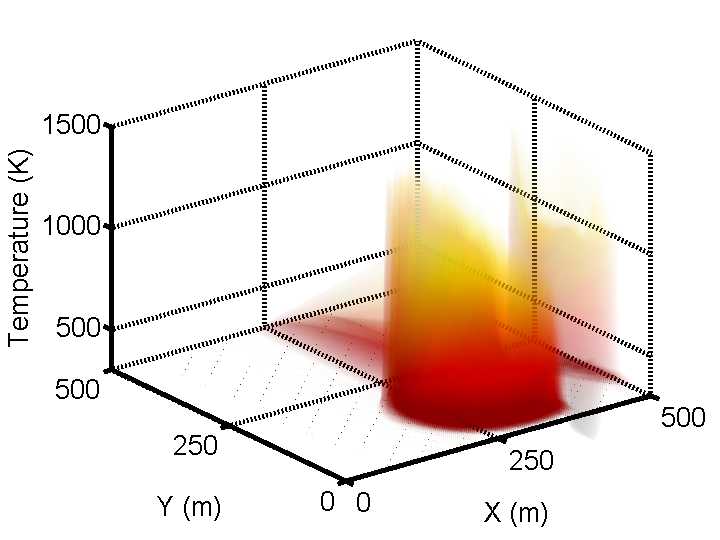}\\
(c) & (d)
\end{tabular}
\newline
\end{center}
\caption{The morphing EnKF applied to the reaction-diffusion model
(\ref{eq:heat}) -- (\ref{eq:reaction}). False color is generated from the
temperature with shading for depth perception. The reference solution (a) is
the simulated data. The initial ensemble was created by a random perturbation
of the comparison solution (b), where the fire was ignited at an intentionally
incorrect location. The standard ENKF panel (c) is the result of data
assimilation of the temperature field after running the model for 500 seconds.
The morphing EnKF panel (d) is the ensemble with the image registration
against the temperature field at the time of ignition, and applied to both the
temperature and the fuel supply. The ensembles have 25 members each and they
are visualized as superposition of transparent images of their members. The
observation function was the temperature field on the whole grid. The standard
EnKF ensembles diverges from the data, while the morphing EnKF ensemble keeps
closer to the data.}%
\label{fig:morph-pde}%
\end{figure}

\begin{figure}[ptb]
\begin{center}%
\begin{tabular}
[c]{cc}%
\includegraphics[width=2.6in]{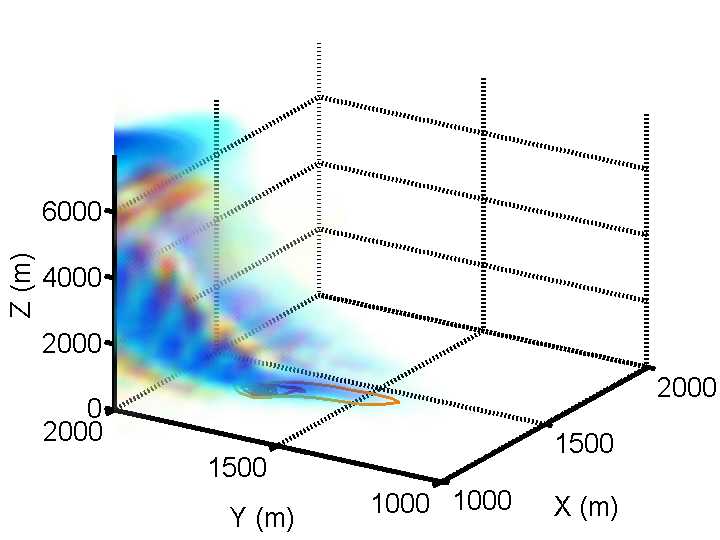} &
\includegraphics[width=2.6in]{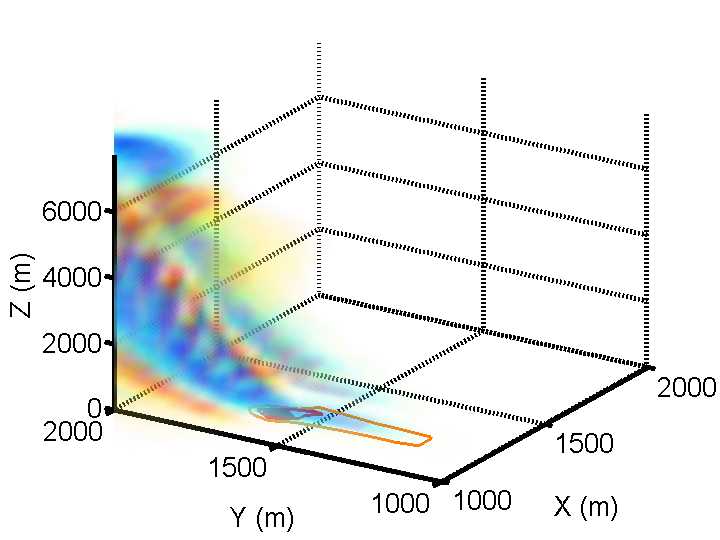}\\
(a) & (b)\\
\includegraphics[width=2.6in]{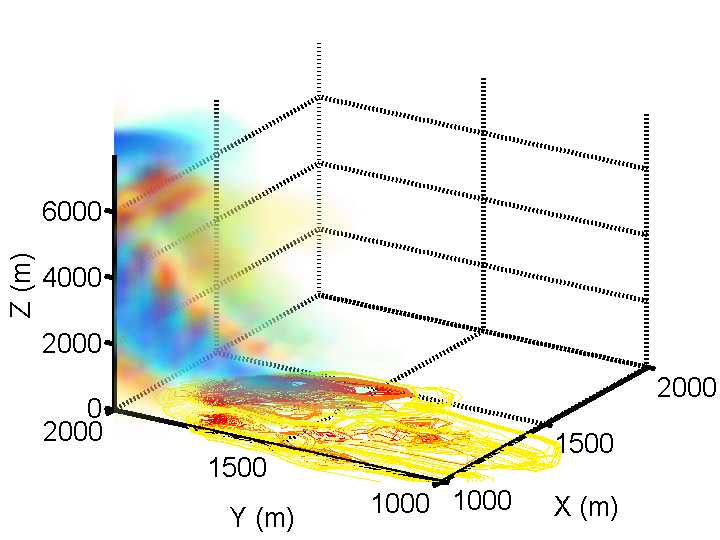} &
\includegraphics[width=2.6in]{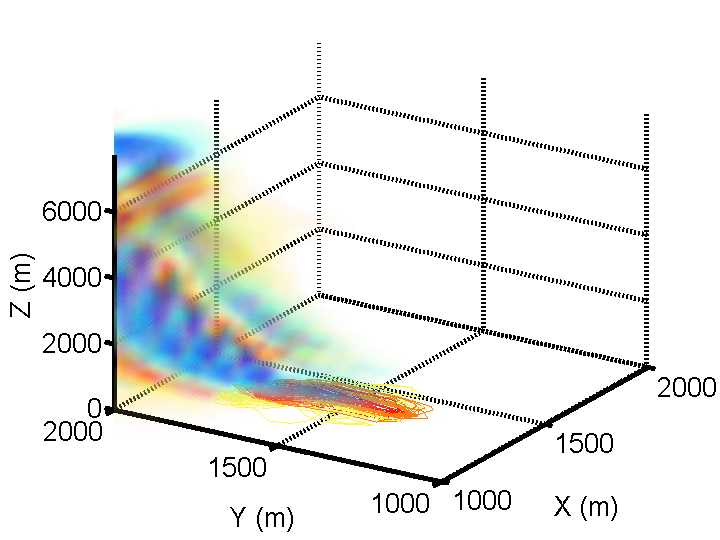}\\
(c) & (d)
\end{tabular}
\newline
\end{center}
\caption{The morphing EnKF applied to the fireline propagation model coupled
with WRF. False color and contour on the horizontal plane is the output heat
flux. The volume shading is the vorticity of the atmosphere where red and blue
shades represent positive and negative vorticity, respectively. A contour plot
of the heat flux due to the fire is shown on the x-y plane. The reference
solution (a) is the simulated data. The initial ensemble was created by a
random perturbation of the comparison solution (b), with the fire ignited at
an intentionally incorrect location. The standard ENKF (c) and the morphing
EnKF (d) were applied after 15 minutes. The ensembles have 25 members each.
Shown are the mean vorticity and the superposition of the heat fluxes in
ensemble members. The registration was done on the atmospheric grid with the
fire heat flux as the observation function. The standard EnKF ensembles
diverges from the data, while the morphing EnKF ensemble keeps closer to the
data.}%
\label{fig:morph-coupled}%
\end{figure}

The morphing EnKF has been demonstrated to be capable of making a large
correction in the state and tracking the data in an accurate and stable manner
for the reaction-diffusion model (Fig. \ref{fig:morph-pde}) as well as for the
level set-based fireline propagation model (Fig. \ref{fig:morph-level}) alone
or coupled with an atmosphere model (Fig. \ref{fig:morph-coupled}).

\section{Conclusion}

Two wildland fire models and methods for assimilating data in those models are
presented. The EnKF is implemented in a distributed-memory high-performance
computing environment. Data assimilation methods are developed combining EnKF
with Tikhonov regularization to avoid nonphysical states and with the ideas of
registration and morphing from image processing to allow large position
corrections. The data assimilation methods can track the data even in the
presence of the large correction, and avoid divergence. The methods can
assimilate gridded data, but the assimilation of station data and steering of
data acquisition is left to future developments.

A semi-empirical fire spread model is implemented by the level set model and
coupled with the Weather Research Forecasting (WRF) model.\ The coupled
fire-atmosphere code is distributed with the WRF download as of Spring 2009,
and it is also available directly from the authors.\newpage

\section{Acknowledgments}

This work is a part of an effort to build a Dynamic Data Driven Application
System (DDDAS) \cite{Darema-2004-DDD} for wildland fire prediction, funded by
the National Science Foundation (NSF). This research was partially supported
by NSF grants CNS-0325314, CNS-0324910, CNS-0719641, DMS-0623983, ATM-0835579,
and ATM-0835598, and by the National Center for Atmospheric Research (NCAR)
Faculty Fellowship.

Computer time on IBM BlueGene/L\ supercomputer was provided by NSF MRI Grants
CNS-0421498, CNS-0420873, and CNS-0420985, NSF sponsorship of the NCAR, the
University of Colorado, and a grant from the IBM Shared University Research
(SUR) program.

The authors would like to thank Craig Douglas, Deng Li, Wei Li, and Adam
Zornes from the University of Kentucky for their contributions to the software
infrastructure that some of the codes used here are built on, John Michalakes
from NCAR for his assistance with WRF, and Ned Patton from NCAR for providing
his prototype code for interfacing an earlier fireline propagation model with WRF.

\newpage
%\bibliographystyle{siam}  % use for arXiv
%\bibliographystyle{IEEEtran}
%\bibliography{csm-fires}

%EndExpansion

%Generated by IEEEtran.bst, version: 1.12 (2007/01/11)

%

%TCIMACRO{\QSubDoc{Include csm-fires-main2}{\include{csm-fires-main2}}}%
%BeginExpansion
%\documentclass[onecolumn,letterpaper,12pt]{IEEEtran}% single spaced, for report
%\usepackage{graphicx,url} % single spaced, for report
% \usepackage{chapterbib}%
%ma double spaced, submit
% double spaced, submit
%\documentclass[onecolumn,letterpaper,12pt]{IEEEtran}% single spaced, for report
% \usepackage{chapterbib}%
%! comment out for report
%! comment out for report

%TCIDATA{OutputFilter=latex2.dll}
%TCIDATA{Version=4.10.0.2363}
%TCIDATA{LaTeXparent=0,0,csm-fires.tex}

\section{Author Information}

Jan Mandel is Professor of Mathematical and Statistical Sciences and Adjunct
Professor of Computer Science and Engineering at the University of Colorado
Denver, Denver, CO, and visiting scientist at the National Center for
Atmospheric Research, Boulder, CO, email \texttt{Jan.Mandel@cudenver.edu}. He
received his PhD in Numerical and Approximate Methods from the Charles
University, Prague, Czech Republic, in 1983. He has developed domain
decomposition and multilevel iterative solvers for large problems of equations
from finite elements. He is interested in high-performance computing, wildfire
modeling, and the development and probabilistic analysis of data assimilation methods.

Jonathan D. Beezley is a PhD student of Applied Mathematics at the University
of Colorado Denver and visiting graduate student at the National Center for
Atmospheric Research, Boulder, CO. He is interested in ensemble filters,
parallel software, and image registration.

Janice L. Coen is a project scientist at the National Center for Atmospheric
Research, Boulder, CO. She received a PhD in Geophysical Sciences from the
University of Chicago in 1992. Her interests are in mesoscale meteorology and
wildland fire behavior. She has developed components of a coupled
atmosphere-fire model and led the implementation of an operational weather
forecasting system for the United Arab Emirates.

Minjeong Kim is a PhD student of Applied Mathematics at the University of
Colorado Denver. She is interested in numerical schemes for level set methods
and in wildfire modeling by partial differential equations.

\newpage

%main text figures

\begin{figure}[ptb]
\begin{center}%
\begin{tabular}
[c]{ccccc}%
\hspace*{-1em}\includegraphics[width=1.4in]{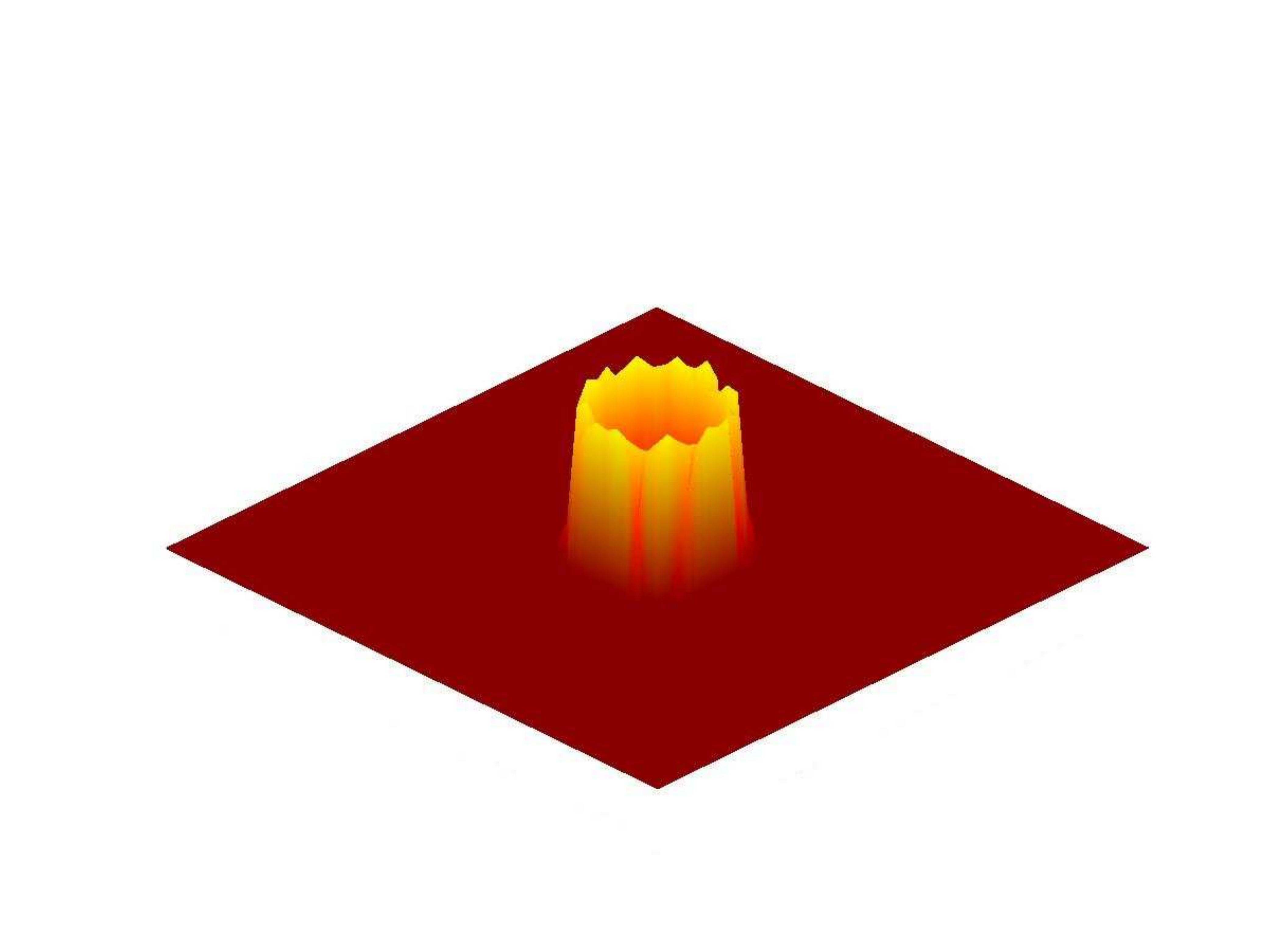}\hspace*{-1em} &
\hspace*{-1em}\includegraphics[width=1.4in]{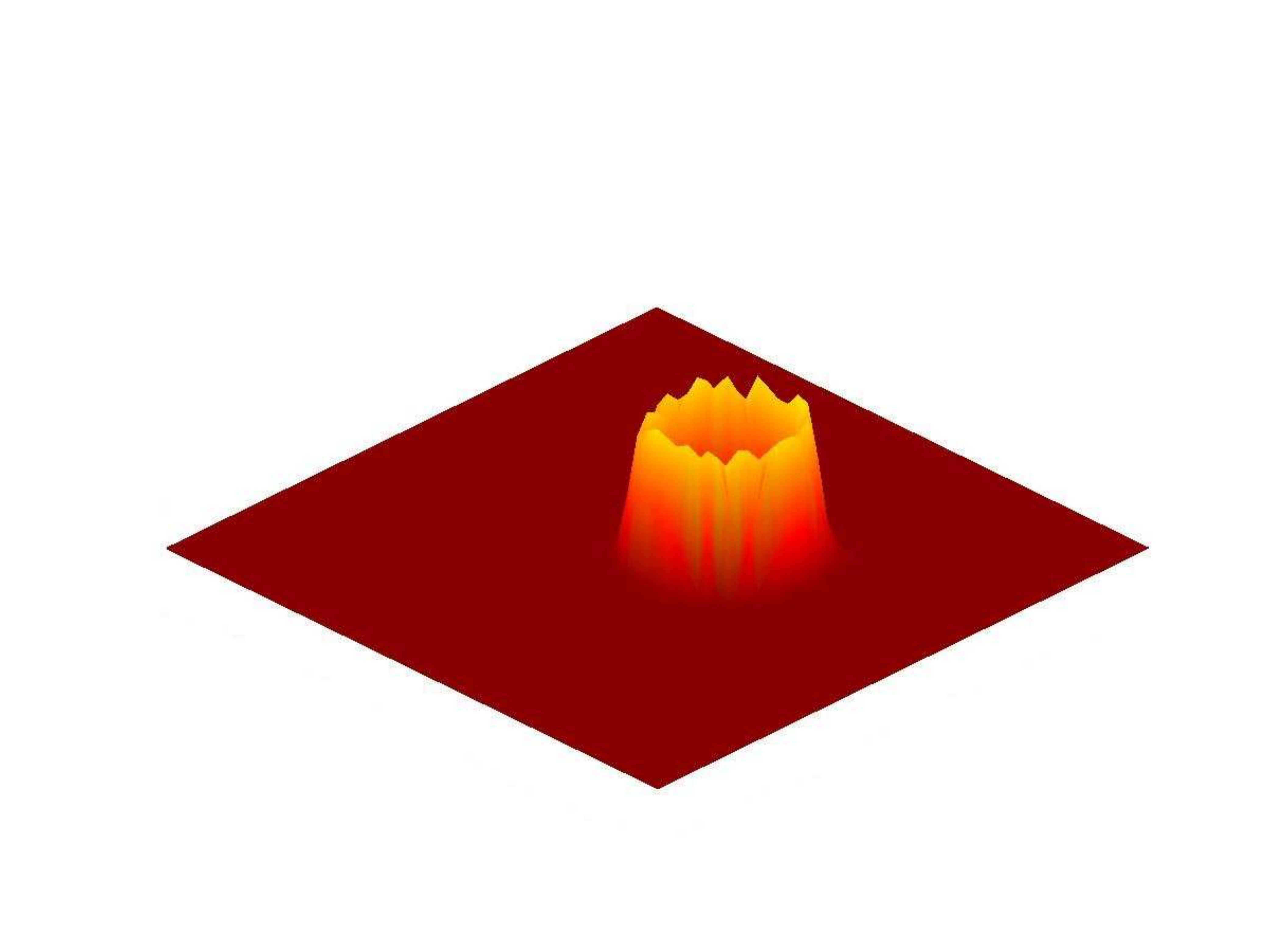}\hspace*{-1em} &
\hspace*{-1em}\includegraphics[width=1.4in]{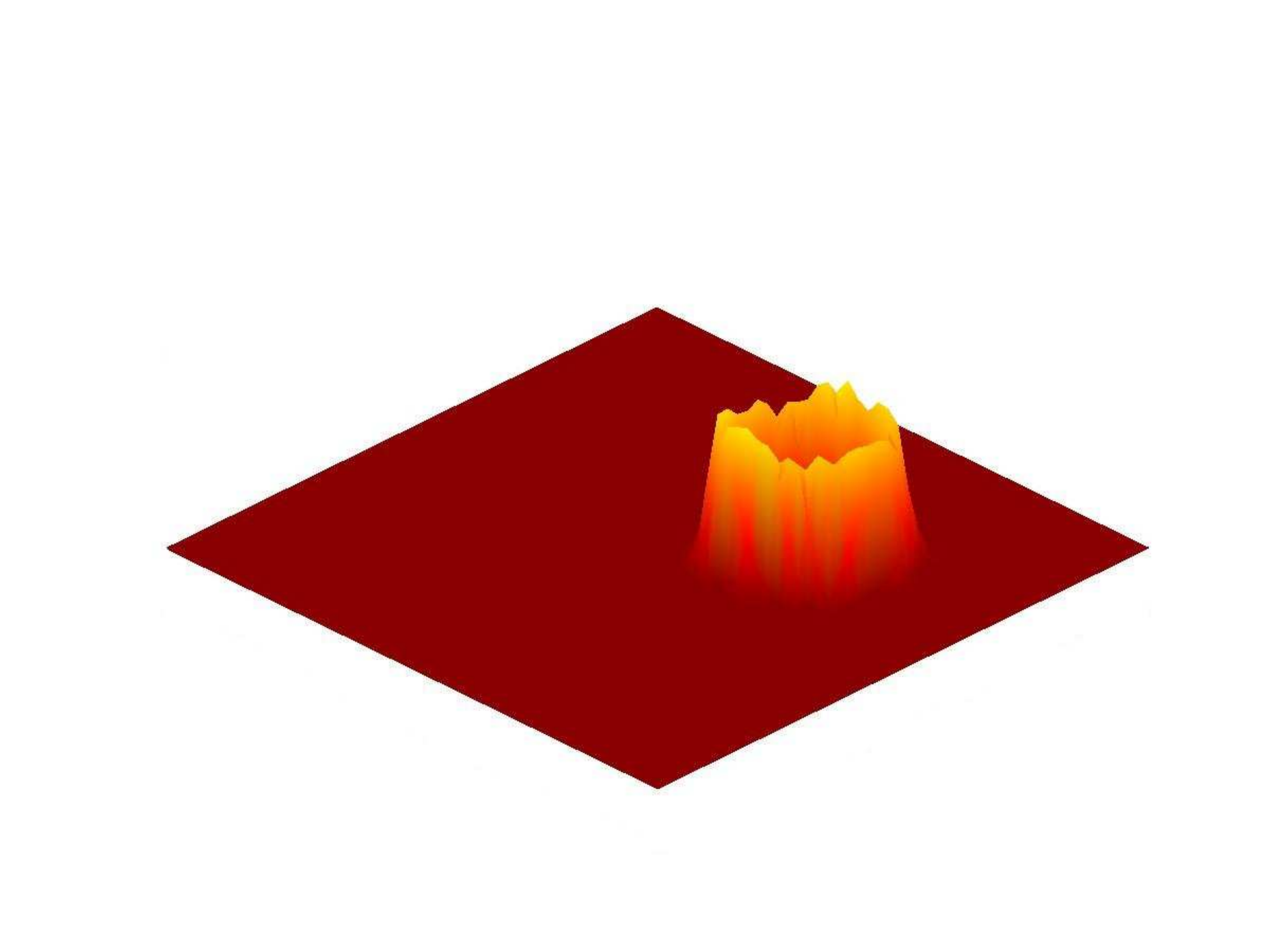}\hspace*{-1em} &
\hspace*{-1em}\includegraphics[width=1.4in]{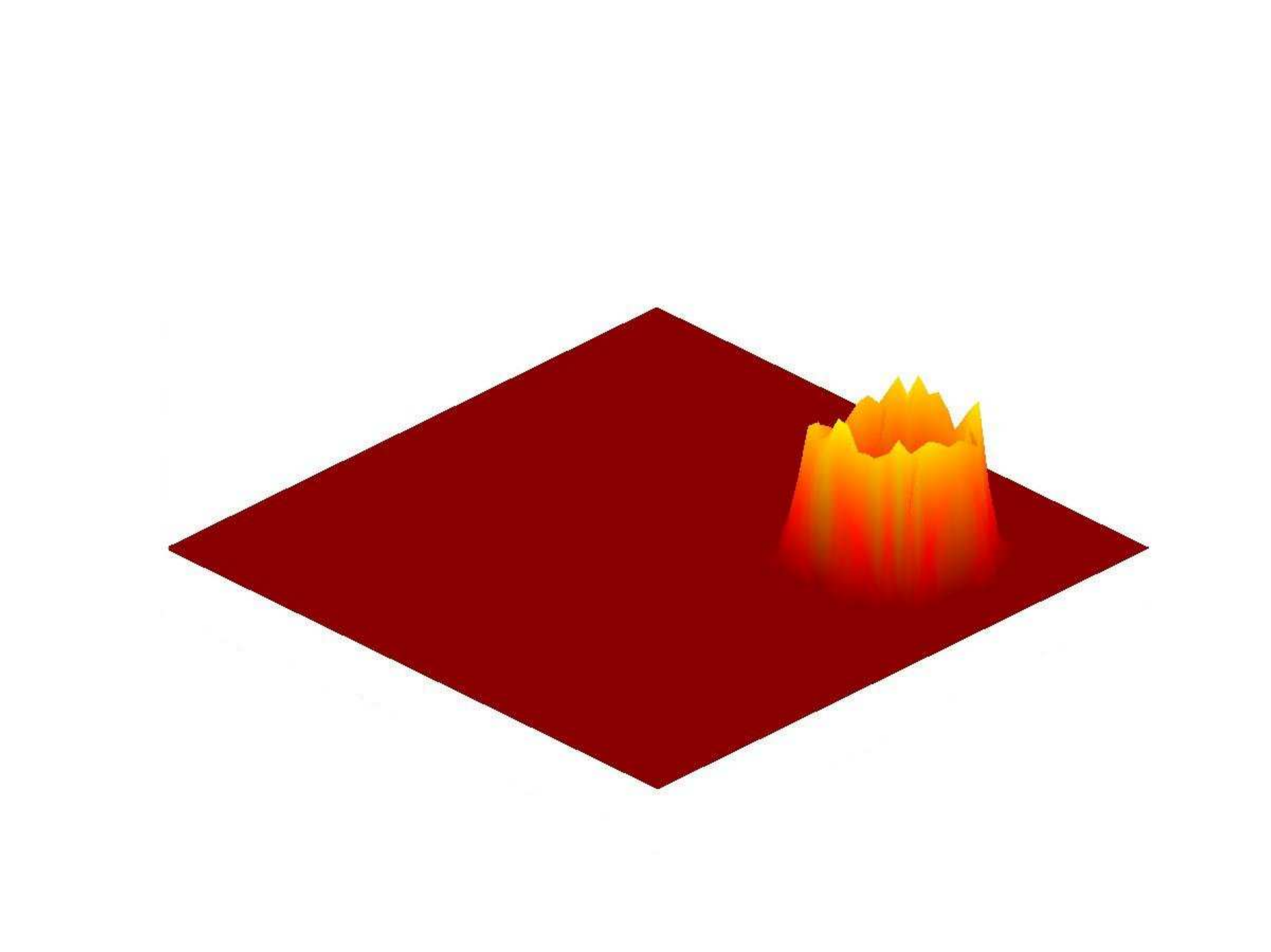}\hspace*{-1em} &
\hspace*{-1em}\includegraphics[width=1.4in]{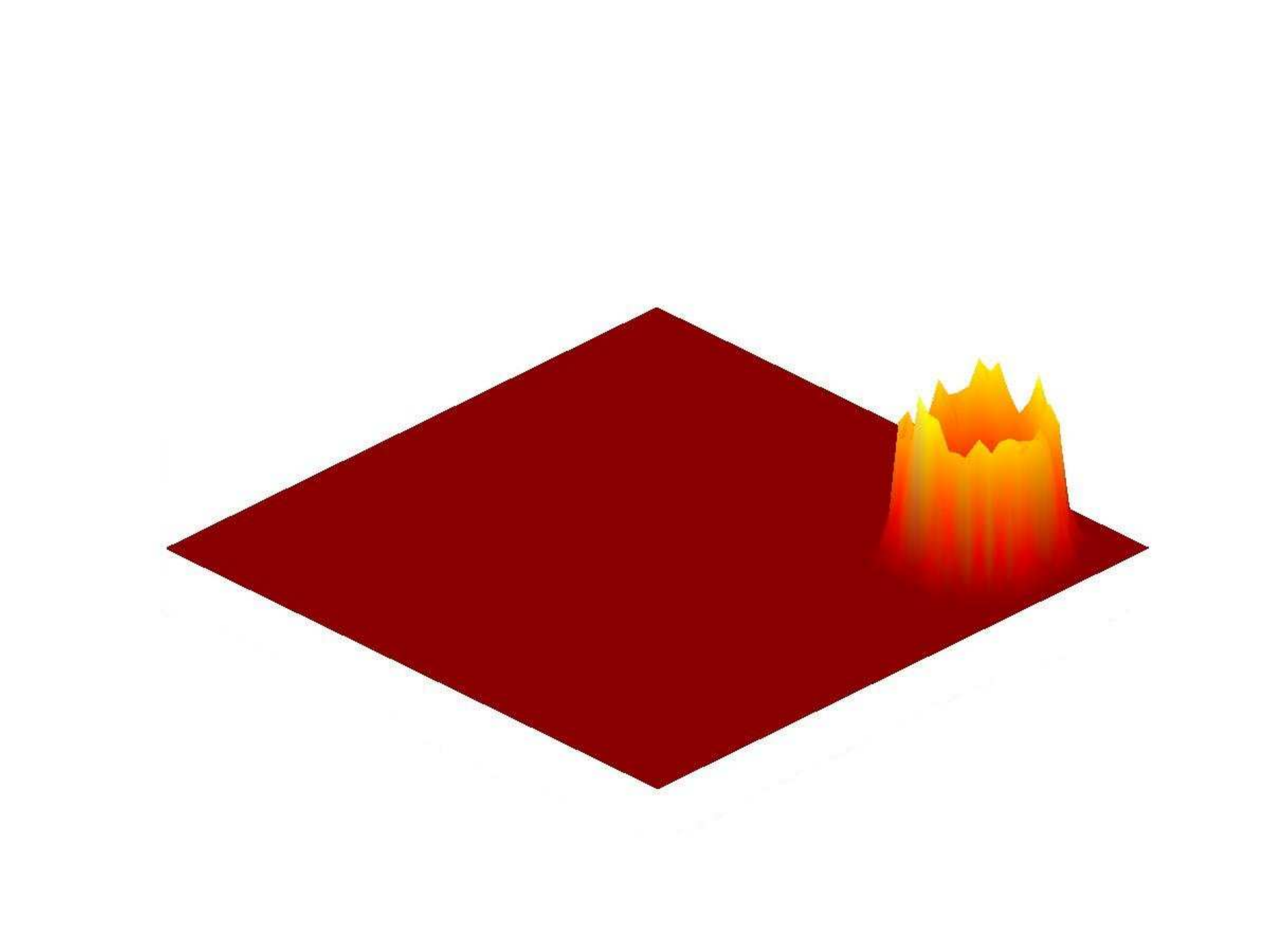}\hspace*{-1em}\\
$\lambda=0$ & $\lambda=0.25$ & $\lambda=0.5$ & $\lambda=0.75$ & $\lambda=1$%
\end{tabular}
\newline
\end{center}
\caption{Morphing of two solutions of a~reaction-diffusion equation system
(\ref{eq:heat}) -- (\ref{eq:reaction}) used in a~wildfire simulation. The
states with $\lambda=0$ and $\lambda=1$ are given. The intermediate states
with $0<\lambda<1$ are created automatically by the morphing algorithm
(\ref{eq:ulambda}). The horizontal plane is the earth surface. The vertical
axis and the color map are the temperature. The morphing algorithm combines
the values as well as the positions. Reproduced from \cite{Beezley-2008-MEK}.}%
\label{fig:morph}%
\end{figure}

\setlength{\unitlength}{1in}

%EndExpansion
%

%TCIMACRO{\QSubDoc{Include csm-fires-modeling}{\include{csm-fires-modeling}}}%
%BeginExpansion
% double spaced, submit
% double spaced, submit
%\documentclass[onecolumn,letterpaper,12pt]{IEEEtran}% single spaced, for report
%\usepackage{graphicx,url} % single spaced, for report
% \usepackage{chapterbib}%
% double spaced, submit
% double spaced, submit
%\documentclass[onecolumn,letterpaper,12pt]{IEEEtran}% single spaced, for report
%\usepackage{graphicx,url} % single spaced, for report
% \usepackage{chapterbib}%

%TCIDATA{OutputFilter=latex2.dll}
%TCIDATA{Version=4.10.0.2363}
%TCIDATA{LaTeXparent=0,0,csm-fires.tex}

\section*{Sidebar - Wildland Fire Modeling}

The purpose of wildland fire modeling is to understand and ultimately predict
fire behavior in order to increase safety of firefighters and the public, to
reduce risk and minimize damage, and to protect ecosystems, watersheds, and
air quality.

\subsection{What is fire modeling?}

The process of fire modeling aims to reproduce and anticipate properties of
fire behavior and fire effects. Fire behavior includes the rate of spread of
the flaming front (the interface between burning and unburned fuel), the heat
release rate which is related to the burning intensity, as well as specific
phenomena such as the bending forward of the fire front, transition from
surface to crown fires, and extreme fire activity such as firewhirls. Fire
effects which include ecological and hydrological effects on the landscape,
such as the percentage of forest fuel consumed in prescribed fires, mortality
suffered by trees, or the quantity of smoke produced for health impacts.

Three environmental factors affect wildland fire behavior: weather, fuel
characteristics, and topography. Fuel factors include the type, moisture,
size, shape, amount, and arrangement. Topography factors include the
orientation toward the sun, the slope, and features such as narrow canyons and
barriers such as creeks, roads, and unburnable fuel. Of the three
environmental factors, weather (including factors such as wind, temperature,
relative humidity, and precipitation) is the most rapidly changing. Weather
phenomena that bring changes such as cold fronts, foehn winds, thunderstorm
downdrafts, sea and land breezes, and diurnal slope winds can be particularly
dangerous, since they can suddenly change the fire's direction and behavior.
Weather also influences wildfires through the other factors of fuel and
topography, by controlling the fuel moisture through precipitation, RH, and
winds, and by complicating the fire-accelerating effect of slopes with
topographically-induced accelerations. These factors are not independent.
Weather and terrain combine to produce topographic effects such as downslope
windstorms (called Santa Annas, foehn winds, East winds, depending on the
geographic location), the weather influences fuel moisture as dry winds or
precipitation increases or decreases evaporation of moisture and therefore the
fuel moisture, and fuel properties may vary with topography as plant density
varies with elevation or aspect with respect to the sun. Although weather can
influence wildfire behavior through many pathways, it has long been recognized
that fires \textquotedblleft create their own weather\textquotedblright. That
is, the heat and moisture created by the fire feed back into the atmosphere,
creating intense winds that drive the fire's behavior, sometimes overwhelming
the effect of ambient winds.

\subsection{Approaches to fire modeling}

Wildland fire models span a vast range of complexity, from simple cause and
effect principles to the most physically complex presenting a difficult
supercomputing challenge that cannot hope to be solved faster than real time.
Conceptual models from experience and intuition from past fires can be used to
anticipate the future. Many semi-empirical fire spread equations as in
\cite{Rothermel-1972-MMP}, \cite{Forestry-1992-DSC}, and \cite{Noble-1980-FMD}
and \cite{Cheney-1993-IFW} for Australasian fuel complexes have been developed
for quick estimation of fundamental parameters of interest such as fire spread
rate, flame length, and fireline intensity of surface fires at a point for
specific fuel complexes, assuming a representative point-location wind and
terrain slope. Based on the work in \cite{Fons-1946-AFS} and
\cite{Emmons-1963-FF} the quasi-steady equilibrium spread rate calculated for
a surface fire on flat ground in no-wind conditions was calibrated using data
of piles of sticks burned in a flame chamber/wind tunnel to represent other
wind and slope conditions for the fuel complexes tested. A simplified
physically-based two-dimensional fire spread model based upon conservation
laws that use radiation as the dominant heat transfer mechanism and
convection, which represents the effect of wind and slope was developed in
\cite{Asensio-2002-WFM}. Two-dimensional fire growth models such as FARSITE
\cite{Finney-1998-FFA} and Prometheus, the Canadian wildland fire growth model
designed to work in Canadian fuel complexes, have been developed that apply
such semi-empirical relationships and others regarding ground-to-crown
transitions to calculate fire spread and other parameters along the surface.
Certain assumptions must be made in models such as FARSITE and Prometheus to
shape the fire growth for example, Prometheus and FARSITE use the Huygens
principle of wave propagation. A set of equations that can be used to
propagate (shape and direction) a fire front using an elliptical shape was
developed in \cite{Richards-1988-NSF}. Although more sophisticated
applications use a three-dimensional numerical weather prediction system to
provide inputs such as wind velocity to one of the fire growth models listed
above, the input was passive and the feedback of the fire upon the atmospheric
wind and humidity are not accounted for.

More complex physical models join computational fluid dynamics models with a
wildland fire component and allow the fire to feed back upon the atmosphere.
These models include NCAR's Coupled Atmosphere-Wildland Fire-Environment
(CAWFE) model developed in \cite{Coen-2005-SBE}, Los Alamos National
Laboratory's FIRETEC developed in \cite{Linn-2002-SWB}, and the WUI (Wildland
Urban Interface) Fire Dynamics Simulator (WFDS) \cite{Mell-2007-PAM} and, to
some degree, the two-dimensional model PIF97
\cite{Dupuy-1999-FSP,Porterie-1998-MMP}. These tools have different emphases
and have been applied to better understand the fundamental aspects of fire
behavior, such as fuel inhomogeneities on fire behavior \cite{Linn-2002-SWB},
feedbacks between the fire and the atmospheric environment as the basis for
the universal fire shape \cite{Coen-2001-CAF,Clark-2004-DCA}, and are
beginning to be applied to wildland urban interface house-to-house fire spread
at the community-scale. The cost of added physical complexity is a
corresponding increase in computational cost, so much so that a full
three-dimensional explicit treatment of combustion in wildland fuels by direct
numerical simulation (DNS) at scales relevant for atmospheric modeling does
not exist, is beyond current supercomputers, and does not currently make sense
to do because of the limited skill of weather models at spatial resolution
under $1$ km. Consequently, even these more complex models parameterize the
fire in some way, for example, \cite{Clark-1996-CAFb,Clark-1996-CAF} use
\cite{Rothermel-1972-MMP} to calculate local fire spread rates using
fire-modified local winds. And, although FIRETEC and WFDS carry prognostic
conservation equations for the reacting fuel and oxygen concentrations, the
computational grid cannot be fine enough to resolve the reaction rate-limiting
mixing of fuel and oxygen, so approximations must be made concerning the
subgrid-scale temperature distribution or the combustion reaction rates
themselves. These models also are too small-scale to interact with a weather
model, so the fluid motions use a computational fluid dynamics model confined
in a box much smaller than the typical wildfire. The limitations on fire
modeling are not entirely computational. At this level, the models encounter
limits in knowledge about the composition of pyrolysis products and reaction
pathways, in addition to gaps in basic understanding about some aspects of
fire behavior such as fire spread in live fuels and surface-to-crown fire transition.

Thus, while more complex models have value in studying fire behavior and
testing fire spread in a range of scenarios, from the application point of
view, FARSITE and Palm-based applications of BEHAVE have shown great utility
as practical in-the-field tools because of their ability to provide estimates
of fire behavior in real time. While the coupled fire-atmosphere models have
the ability to incorporate the ability of the fire to affect its own local
weather, and model many aspects of the explosive, unsteady nature of fires
that cannot be incorporated in current tools, it remains a challenge to apply
these more complex models in a faster-than-real-time operational environment.
Also, although they have reached a certain degree of realism when simulating
specific natural fires, they must yet address issues such as identifying what
specific, relevant operational information they could provide beyond current
tools, how the simulation time could fit the operational time frame for
decisions (therefore, the simulation must run substantially faster than real
time), what temporal and spatial resolution must be used by the model, and how
they estimate the inherent uncertainty in numerical weather prediction in
their forecast. These operational constraints must be used to steer model development.

%\bibliographystyle{IEEEtran}
%\bibliography{csm-fires}

%EndExpansion

%Generated by IEEEtran.bst, version: 1.12 (2007/01/11)

%

%TCIMACRO{\QSubDoc{Include csm-fires-levelset}{\include{csm-fires-levelset}}}%
%BeginExpansion
% double spaced, submit
% double spaced, submit
%\documentclass[onecolumn,letterpaper,12pt]{IEEEtran}% single spaced, for report
%\usepackage{graphicx,url} % single spaced, for report
% \usepackage{chapterbib}%
% double spaced, submit
% double spaced, submit
%\documentclass[onecolumn,letterpaper,12pt]{IEEEtran}% single spaced, for report
%\usepackage{graphicx,url} % single spaced, for report
% \usepackage{chapterbib}%
% double spaced, submit
% double spaced, submit
%\documentclass[onecolumn,letterpaper,12pt]{IEEEtran}% single spaced, for report
%\usepackage{graphicx,url} % single spaced, for report
% \usepackage{chapterbib}%
% double spaced, submit
% double spaced, submit
%\documentclass[onecolumn,letterpaper,12pt]{IEEEtran}% single spaced, for report
%\usepackage{graphicx,url} % single spaced, for report
% \usepackage{chapterbib}%
%ma double spaced, submit
% double spaced, submit
%\documentclass[onecolumn,letterpaper,12pt]{IEEEtran}% single spaced, for report
%\usepackage{graphicx,url} % single spaced, for report
% \usepackage{chapterbib}%
%ma double spaced, submit
% double spaced, submit
%\documentclass[onecolumn,letterpaper,12pt]{IEEEtran}% single spaced, for report
%\usepackage{graphicx,url} % single spaced, for report
% \usepackage{chapterbib}%

%TCIDATA{OutputFilter=latex2.dll}
%TCIDATA{Version=4.10.0.2363}
%TCIDATA{LaTeXparent=0,0,csm-fires.tex}

\section*{Sidebar - Level Set-Based Wildland Fire Model}

\subsection{Curve propagation by the level set method}

Consider a curve $\Gamma=\Gamma\left(  t\right)  $ in $\mathbb{R}^{2}$
evolving with the speed $S\left(  x,t\right)  $ in the normal direction. The
level set method represents the curve as the level set%
\[
\Gamma=\left\{  x:\psi\left(  x,t\right)  =0\right\}
\]
of a function $\psi$, called a level set function, and converts the evolution
of the curve into a differential equation for the level set function, which is
then solved numerically. In applications, the curve $\Gamma$ is closed, and we
choose that the inside of the curve is given by $\psi<0$. Because $\psi=0$ on
$\Gamma$, the component of $\nabla\psi$ in the tangent direction is zero, and
so $\nabla\psi$ has the normal direction to $\Gamma$. Thus, we can also use
the level set function $\psi$ to compute the normal to the curve $\Gamma$ by%
\[
\overrightarrow{n}=\frac{\nabla\psi}{\left\Vert \nabla\psi\right\Vert },
\]
where $\left\Vert \cdot\right\Vert $ is the Euclidean norm.

To derive an equation to evolve the level set function, consider a point
$x=x\left(  t\right)  \in\Gamma\left(  t\right)  $. Then $\psi\left(  x\left(
t\right)  ,t\right)  =0$, and from the chain rule

%\[%
\begin{align*}
0  &  =\frac{d}{dt}\psi\left(  x,t\right) \\
&  =\frac{\partial\psi}{\partial t}+\underbrace{\nabla\psi}_{\left\Vert
\nabla\psi\right\Vert \overrightarrow{n}}\cdot\frac{\partial x}{\partial t}\\
&  =\frac{\partial\psi}{\partial t}+\left\Vert \nabla\psi\right\Vert
\underbrace{\overrightarrow{n}\cdot\frac{\partial x}{\partial t}}_{S\left(
x\right)  }.
\end{align*}
%\]
Thus, the level set function is governed by the partial differential equation,
called the level set equation \cite{Osher-1988-FPC} \cite[Ch. 1]%
{Sethian-1999-LSM}%

\[
\frac{\partial\psi}{\partial t}+S\left(  x\right)  \left\Vert \nabla
\psi\right\Vert =0.
\]
Note that the level set equation requires propagation speed $S\left(
x\right)  $ to be defined at all points $x$, not just on the curve $\Gamma$,
even if the curve propagation itself depends only on the speed $S$ given on
$\Gamma$.

One example of the level set function is the signed distance function%
\[
\psi\left(  x\right)  =\pm\operatorname*{dist}(x,\Gamma),
\]
with $-$ inside and $+$ outside of $\Gamma$. The signed distance function is
characterized by $\left\Vert \nabla\psi\right\Vert =1$. Ideally, $\psi$ should
be equal or close to the signed distance function for numerical stability, but
this property is in general not preserved when advancing the level set
equation in time.

Since in the wildland fire model here always $S\geq0$, the level set function
does not increase with time and the fire area cannot decrease (Fig.
\ref{fig:prop}).

\subsection{Numerical solution of the level set equation}

The spread rate $S$ is defined by the Rothermel formula (\ref{eq:Rothermel})
at every point, not only the fireline. Thus, the fuel data and wind over the
whole domain can be used to propagate the level set function, avoiding the
problem how to extend the speed from the fireline. The level set equation is
discretized on a rectangular grid rectangular mesh with spacing $\left[
\triangle x_{1},\triangle x_{2}\right]  $. To advance the model in time, the
Runge-Kutta method of order $2$ (Heun's method) is used,
\[
\psi^{n+1/2}=\psi^{n}+\Delta tF\left(  \psi^{n}\right)  ,\quad\psi^{n+1}%
=\psi^{n}+\Delta t\left(  \left(  1- \alpha\right)  F\left(  \psi^{n}\right)  +\alpha
F\left(  \psi^{n+1/2}\right)  \right)  ,\quad\alpha=\frac{1}{2}.
\]
The right-hand side $F$ is a discretization of the term $-S\left\Vert
\nabla\psi\right\Vert $ with upwinding and artificial viscosity,%
\[
F\left(  \psi\right)  =-S\left(  \overrightarrow{v}\cdot\overrightarrow
{n},\nabla z\cdot\overrightarrow{n}\right)  \left\Vert \overline{\nabla}%
\psi\right\Vert +\varepsilon\triangle\psi,\quad\overrightarrow{n}=\nabla
\psi/\left\Vert \nabla\psi\right\Vert ,
\]
where $\overline{\nabla}\psi=\ \left[  \overline{\nabla}_{1}\psi
,\overline{\nabla}_{2}\psi\right]  $ is the upwinded approximation of the
gradient of $\psi$ by the Godunov method \cite[p. 58]{Osher-2003-LSM},
\[
\overline{\nabla}_{1}\psi=\left\{
\begin{tabular}
[c]{ll}%
$\nabla_{1}^{-}\psi$, & if $\nabla_{1}^{-}\psi\geq0$ and $\nabla_{1}^{-}%
\psi+\nabla_{1}^{+}\psi\geq0,$\\
$\nabla_{1}^{+}\psi$, & if $\nabla_{1}^{+}\psi\leq0$ and $\nabla_{1}^{-}%
\psi+\nabla_{1}^{+}\psi<0$,\\
$0$, & otherwise,
\end{tabular}
\ \ \right.
\]
where $\nabla_{1}^{+}\psi$ and $\nabla_{1}^{-}\psi$ are the right and the left
one-sided numerical derivatives%
\begin{align*}
\nabla_{1}^{+}\psi\left(  x_{1},x_{2}\right)   &  =\frac{\psi\left(
x_{1}+\triangle x_{1},x_{2}\right)  -\psi\left(  x_{1},x_{2}\right)
}{\triangle x_{1}},\\
\nabla_{1}^{-}\psi\left(  x_{1},x_{2}\right)   &  =\frac{\psi\left(
x_{1},x_{2}\right)  -\psi\left(  x_{1}-\triangle x_{1},x_{2}\right)
}{\triangle x_{1}},
\end{align*}
and similarly for $\nabla_{2}^{+}\psi$ and $\nabla_{2}^{-}\psi$. Further,
\[
\nabla\psi=\left[  \frac{\nabla_{1}^{+}\psi+\nabla_{1}^{-}\psi}{2}%
,\frac{\nabla_{2}^{+}\psi+\nabla_{2}^{-}\psi}{2}\right]
\]
is the gradient by central differences, $\varepsilon$ is the scale-free
artificial viscosity ($\varepsilon=0.4$ in the computations here), and%
\[
\triangle\psi=\left(  \nabla_{1}^{+}\psi-\nabla_{1}^{-}\psi+\nabla_{2}^{+}%
\psi-\nabla_{2}^{-}\psi\right)
\]
is the scaled five-point Laplacian of $\psi$. The wind velocity
$\overrightarrow{v}\left(  x\right)  $ can be evaluated at $x-d\overrightarrow
{n}$ instead, i.e., at the fireline, the winds are taken from the distance $d$
behind the fireline. 

The use of a numerically stable scheme that includes upwinding, such as the
Godunov scheme above in the norm of the gradient, is required; when the
gradient there is approximated by standard central differences, the numerical
method fails quickly. However, it seems better to use central differences in
the computation of the normal $\overrightarrow{n}$.

The reason for using Heun's method is not accuracy but conservation. In
numerical computations, the second time derivative of the level set function
tends be negative, $\partial^{2}\psi/\partial^{2}t<0$, and so the explicit
Euler method ($\alpha=0$ above) tends to systematically overestimate $\psi$
and thus fire propagation slows down or stops altogether. Similarly, the
choice $\alpha=1$ tends to systematically underestimate $\psi$. Heun's method
($\alpha=1/2$) as the average of the two behaves reasonably well.

\subsection{Boundary conditions}

Before computing the one-sided differences up to the boundary, the level set
function is extrapolated to one layer of nodes beyond the boundary. However,
the extrapolation is not allowed to decrease the value of the level set
function under the value at the boundary. For example, when $\left(
n,k\right)  $ is the last node in the domain in the direction $x_{1},$ the
extrapolation
\[
\psi_{n+1,k}=\max\left\{  \psi_{nk}+\left(  \psi_{nk}-\psi_{n-1,k}\right)
,\psi_{nk},\psi_{n-1,k}\right\}  ,
\]
is used, and similarly in other cases. This modification of the finite
difference method serves to avoid numerical instabilities at the boundary. The
extrapolation at the boundary effectively implements a free boundary
condition. Without the stabilization, a decrease of the level set function at
boundary node, e.g., due to nonhomogeneity of the fuel, would be 
amplified by the extrapolation, and, in combination with
the effect of the Godunov formula, might cause large negative spikes to develop at the boundary.

\subsection{Updating ignition time}

The ignition time $t_{i}$ in the strip that the fire has moved over in one
timestep is computed by linear interpolation from the level set function. Suppose
that the point $x$ is not burning at the time $t$ but it is burning at the
time $t+\triangle t$, that is, $\psi\left(  x,t\right)  >0$ and $\psi\left(
x,t+\triangle t\right)  \leq0$. The ignition time $t_{i}\left(  x\right)  $ at
the point $x$ satisfies $\psi\left(  x,t_{i}\right) =0 $. Approximating $\psi$
linearly in $t$, we have%
\[
\frac{\psi\left(  x,t\right)  }{t_{i}\left(  x\right)  -t}\approx-\frac
{\psi\left(  x,t+\triangle t\right)  }{t_{i} \left(  x\right)-\left(  t+\triangle t\right)  },
\]
which gives%
\[
t_{i}(x)\approx t+\frac{\psi\left(  x,t\right)  \triangle t}{\psi\left(
x,t\right)  -\psi\left(  x,t+\triangle t\right)  }.
\]

\subsection{Computation of fuel burned}

The fuel burned and thus the heat generated are then computed by numerical
quadrature over each fire mesh cell from the postulated exponential fuel decay
(\ref{eq:burning}). This is a challenging task because the numerical
quadrature needs to be able to handle two extreme cases. When the fuel burns
slowly relative to the time the fireline takes to cross the cell, the
integrand is well approximated by a linear function and the proportion of the
fuel burned is related to the volume of the burning area. On the other hand,
when the fuel burns quickly relative to the time step and the fireline moves
slowly relative to the mesh step size, the fuel burned in one timestep is
related to the area of a strip through which the fireline travels during the
time step, and asymptotically to the length of the fireline in the cell.

The level set function $\psi$ and the ignition time $t_{i}$ are interpolated
to the corners of subcells that form at least twice finer grid. On each
subcell $C$, the fuel fraction burned%
\[
\frac{1}{\operatorname*{area}\left(  C\right)  }%
%TCIMACRO{\diint \limits_{x\in C:\psi\left(  x\right)  \leq0}}%
%BeginExpansion
{\displaystyle\iint\limits_{x\in C:\psi\left(  x\right)  \leq0}}
%EndExpansion
\left(  1-e^{-\frac{t-t_{i}}{T}}\right)  dx_{1}dx_{2}%
\]
is approximated from the values $\psi$ and $t_{i}$ at the corners of the
subcell. The fraction of the cell that is burning is approximated by
\[
\operatorname*{area}\left\{  x\in C:\psi\left(  x\right)  \leq0\right\}
\approx\beta=\frac{1}{2}\left(  1-\frac{\sum_{k=1}^{4}\psi_{k}}{\sum_{k=1}%
^{4}\left\vert \psi_{k}\right\vert }\right)  ,
\]
where $\psi_{1},\ldots,\psi_{4}$ are the values of the level set function at
the corners of the subcell. The approximation is exact in several important
cases:\ No part of the cell is on fire, all $\psi_{k}\geq0$ and at least one
$\psi_{k}>0$, then $\beta=0$; the whole cell is on fire, all $\psi_{k}\leq0$
and at least one $\psi_{k}<0$, then $\beta=1$; or the values $\psi_{k}$ define
a linear function and the fireline crosses the cell diagonally or it is
aligned with one coordinate direction.

Next, replace the time from ignition on the subcell corners by zero whenever
the level set function is positive (then the corner cannot be on fire), and
compute the approximate fraction of the fuel burned as $\beta\left(
1-e^{-t_{a}/W}\right)  t_{a}$, where $W$ is the time constant of the fuel in
the cell and $t_{a}$ is the average of the modified time from ignition on the
subcell corners. This computation is a second-order quadrature formula when
the whole cell is burning, accurate when no part of the cell is burning, and
it provides a natural transition between the two. Also, the calculation is
accurate asymptotically when the fuel burns slowly and the approximation
$\beta$ of the burning area is exact. However, a rigorous investigation of the
accuracy of the quadrature is beyond the scope of this paper.

\subsection{Ignition}

The model is initialized with no fire by choosing the level set function
$\psi\left(  x\right)  =1$. If a given ignition time $t_{0}$ falls within the
time step, then the level set function is multiplied by a function that is
negative in the fire region, before entering the time step. For ignition in a
circle of radius $r$ around a point $y$, the level set function $\psi\left(
x\right)  $ is replaced by%
\[
\left(  \left\Vert x-y\right\Vert -r\right)  \psi\left(  x\right)  ,
\]
\newline Ignition within radius $r$ of a more general object, such as a line
$L$, is implemented by replacing the level set function by
\[
\left(  d\left(  x,L\right)  -r\right)  \psi\left(  x\right)
\]
where $d\left(  x,L\right)  $ is the Euclidean distance of the point $x$ from
$L$. To avoid discretization artefacts, the ignition radius should be several
mesh sizes large.

The ignition time on newly ignited nodes is currently set to the current time
$t_{0}$. Alternatively, the ignition time could be initialized to%
\[
t_{i}\left(  x\right)  =t_{0}+R_{0}\left(  \left\Vert x-x_{0}\right\Vert
-s\right)  ,\quad\left\Vert x-x_{0}\right\Vert \leq s,
\]
which gives the time of ignition $t_{0}$ at the boundary of the circle and
$t_{0}-R_{0}s$ at the center.

\subsection{Related work}

One problem with level set methods is that they do not guarantee the
\emph{a-priori} conservation of the shape or volume of the level set even in a
uniform constant velocity field. Because of the numerical dispersion inherent
in upwinding schemes, the shape of level set function flattens up and the
level set may either grow too much or decrease and eventually vanish. Also,
details of the level set get smeared, and the level set gets seriously
distorted, e.g., in a rotational velocity field. For this reason, high-order
essentially non-oscillatory (ENO) finite differences in space and high-order
Runge-Kutta methods in time are recommended \cite{Osher-1988-FPC} \cite[Ch.
1]{Sethian-1999-LSM} and indeed required in, e.g., the code from
\cite{Fendell-2001-WFS}. In addition, sophisticated approaches have been
developed to improve the conservation of the level set, such as the hybrid
particle level set method \cite{Enright-2005-FAS}, quadrature-free
discontinuous Galerkin method \cite{Marchandise-2006-QDG}, and flux-based
level set method \cite{Frolkovic-2007-FLS}.

However, the condition that the fire always spreads forward by a minimum
backing rate $B_{0}\geq0$, and thus the level set function can only decrease,
appears to prevent numerical artefacts in the fire model. The much simpler
Godunov upwinding and the second-order Runge-Kutta method along with the
stabilizations described above are sufficient in the computations reported here.

In \cite{Mallet-2007-MWF}, the narrow band level set method with velocity
extension from the nearest point on the fireline was used to implement the
fireline propagation model from \cite{Fendell-2001-WFS}. The model postulates
different spread rates at the head and the flanks of a fire and it does not
consider the fuel balance or a coupling with the atmosphere. The narrow band
level set method maintains only the values of the level set function in a band
along the fireline. It can save a large number of operations and it may reduce
the tendency of level set methods to distort the fire area, because the pieces
of the level set function emanating from different parts of the fireline do
not need to be reconciled globally. However, a narrow band level set method
does not seem to be as suitable for data assimilation, it does not fit
naturally in the framework of geophysical codes, which operate on rectangular
arrays, and the level set method as described above seems to be satisfactory.

%\bibliographystyle{IEEEtran}
%\bibliography{csm-fires}

%EndExpansion

%Generated by IEEEtran.bst, version: 1.12 (2007/01/11)

%

%TCIMACRO{\QSubDoc{Include csm-fires-registration}{\include
%{csm-fires-registration}}}%
%BeginExpansion
% double spaced, submit
% double spaced, submit
%\documentclass[onecolumn,letterpaper,12pt]{IEEEtran}% single spaced, for report
%\usepackage{graphicx,url} % single spaced, for report
% \usepackage{chapterbib}%
% double spaced, submit
% double spaced, submit
%\documentclass[onecolumn,letterpaper,12pt]{IEEEtran}% single spaced, for report
%\usepackage{graphicx,url} % single spaced, for report
% \usepackage{chapterbib}%

%TCIDATA{OutputFilter=latex2.dll}
%TCIDATA{Version=4.10.0.2363}
%TCIDATA{LaTeXparent=0,0,csm-fires.tex}

\section*{Sidebar - Image Registration}

Moving and stretching one given image to become another given image is known
in image processing as registration \cite{Brown-1992-SIR}. The procedure
outlined here can do that fully automatically, without a human intervention.

Given function $u$ and $v$ on a domain $D$, the registration problem can be
formulated as
\[
v\approx u\circ\left(  I+T\right)  \text{ on }D.
\]
At the same time, we wish that $T$ and $\bigtriangledown T$ are not
unnecessarily large. These requirements naturally lead to a construction of
the mapping $T$ by optimization%
\[
J(T)=\Vert v-u\circ\left(  I+T\right)  \Vert+C_{1}\Vert T\Vert+C_{2}%
\Vert\bigtriangledown T\Vert\rightarrow\min_{T},
\]
where $\Vert\cdot\Vert$ are suitable norms. We choose the $L^{1}\left(
D\right)  $ norms, $\Vert v\Vert=\int_{D}\left\vert v\right\vert dx$,
evaluated by simple numerical integration. To define $\Vert T\Vert$ and
$\Vert\bigtriangledown T\Vert$, we add the norms of the entries of the vector
function $T$ and the matrix function $\bigtriangledown T$.

The optimization formulation tries to balance the conflicting objectives of
good approximation by the registered image, and as small and smooth warping as
possible. The objective function $J(T)$ is in general not a convex function of
$T$, and there can be many local minima. For example, a local minimum may
occur when some small part of $u$ and $v$ matches, while the overall match is
still not good.

The optimization problem for the mapping $T$ is solved by the algorithm from
\cite{Gao-1998-WMA} with some modifications. The method proceeds by building
the mapping $T$ on a~nested hierarchy of meshes, starting with a coarse mesh
that divides the domain $D$ into just four rectangular cells. The mapping $T$
is interpreted as the movement of grid nodes, and the nodes are moved one at a
time to minimize a coarse version of $J\left(  T\right)  $. For each node, a
mesh of a number of possible locations is constructed, all locations on the
mesh are tested, then the location with the smallest value of $J(T)$ is
selected and further improved by a standard descent method. In the coarse
version of $J\left(  T\right)  $, the mapping $T$ is interpolated from the
nodes of the current grid, and the functions $u$ and $v$ are smoothed by a
convolution to smear fine features and allow a more global match. The
resulting mapping is then interpolated to a mesh that is twice finer, and the
process continues. See \cite{Beezley-2008-MEK} for further details.

%\bibliographystyle{IEEEtran}
%\bibliography{csm-fires}

%EndExpansion

%Generated by IEEEtran.bst, version: 1.12 (2007/01/11)

\newpage

%sidebar figures

\renewcommand{\thefigure}{S\arabic{figure}} \setcounter{figure}{0}

\begin{figure}[ptb]
\begin{center}
\includegraphics[width=4in]{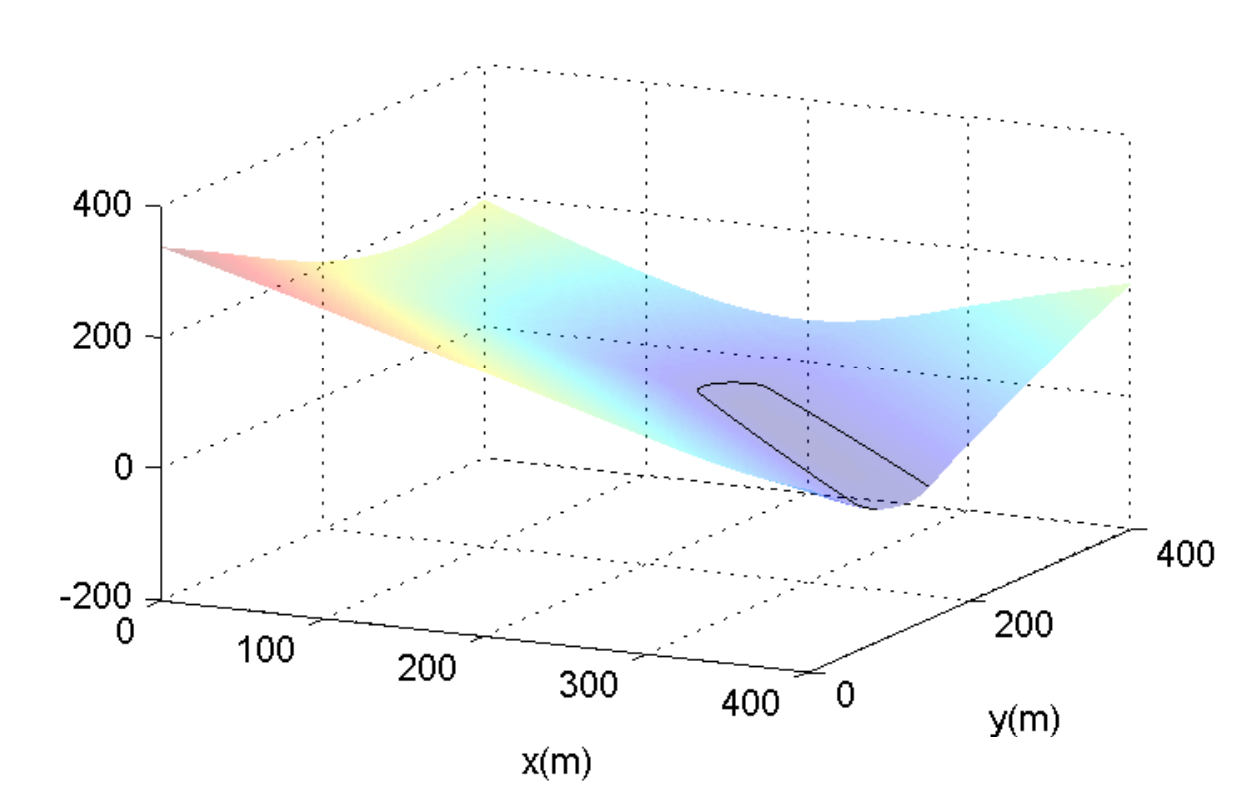}
\end{center}
\caption{The fireline and level set function for a circular fire propagating
in constant wind. The fire is in the area where the level set function is
negative. The black contour is the fireline. The fire starts at the back of
the domain and propagates diagonally until it leaves the domain.}%
\label{fig:prop}%
\end{figure}\ 
\end{document}